\documentclass[12pt, draftclsnofoot, onecolumn]{IEEEtran}
\usepackage{amsmath,amssymb,amsfonts,amsthm}
\usepackage{algorithm,algorithmic}
\usepackage{graphicx,xcolor}
\usepackage{setspace}
\usepackage[caption=false]{subfig}

\def\ie{{\it i.e.,\ \/}}

\def\eg{{\it e.g.,\ \/}}
\DeclareMathOperator{\tr}{tr}

\DeclareMathOperator{\blkdiag}{blkdiag}
\theoremstyle{definition}
\newtheorem{theorem}{Theorem}
\newtheorem{lemma}{Lemma}
\newtheorem{corollary}{Corollary}
\newtheorem{remark}{Remark}
\newtheorem{assumption}{Assumption}
\makeatletter
\allowdisplaybreaks
\begin{document}

\title{
\begin{huge} Periodic Updates for Constrained OCO with Application to Large-Scale Multi-Antenna Systems \end{huge}}

\author{
Juncheng Wang, \IEEEmembership{Student Member, IEEE}, 
Min Dong, \IEEEmembership{Senior Member, IEEE},\\
Ben Liang, \IEEEmembership{Fellow, IEEE},        
and Gary Boudreau, \IEEEmembership{Senior Member, IEEE}
        
\thanks{ 
J.~Wang and B.~Liang are with the University of Toronto (e-mail: \{jcwang, liang\}@ece.utoronto.ca). M.~Dong is with the Ontario Tech University (e-mail: min.dong@ontariotechu.ca). G.~Boudreau is with Ericsson Canada (e-mail: gary.boudreau@ericsson.com). This work has been funded in part by Ericsson Canada and by the Natural Sciences and Engineering Research Council (NSERC) of Canada. A preliminary version of this work has appeared in IEEE SPAWC \cite{SPAWC20}.

}}

\maketitle

\begin{abstract}
In many dynamic systems, decisions on system
operation are updated over time, and the decision maker requires an online
learning approach to optimize its strategy in response to the changing environment.
When the loss and constraint functions are convex, this belongs to the general
family of online convex optimization (OCO). In existing OCO works, the environment
is assumed to vary in a time-slotted fashion, while the decisions are updated
at each time slot. However, many wireless communication systems permit only
periodic decision updates, \ie each decision is fixed over multiple time
slots, while the environment changes between the decision epochs. The standard OCO model is inadequate for these systems. Therefore, in this
work, we consider periodic decision updates for OCO. We aim to minimize the
accumulation of time-varying convex loss functions, subject to both short-term
and long-term constraints. Information about the loss functions within the
current update period may be incomplete and is revealed to the decision maker
only after the decision is made. We propose an efficient algorithm, termed
Periodic Queueing and Gradient Aggregation (PQGA), which employs novel periodic
queues together with possibly multi-step aggregated gradient descent to update
the decisions over time. We derive upper bounds on the dynamic regret, static regret, and constraint violation of PQGA. As an example application, we study the performance of PQGA in a large-scale multi-antenna system shared by multiple wireless service providers. Simulation results show that PQGA converges fast and substantially outperforms the known best alternative.\end{abstract}

\begin{IEEEkeywords}
Online convex optimization, long-term constraint, periodic updates,
massive MIMO, wireless network virtualization.
\end{IEEEkeywords}

\maketitle


\section{Introduction}
\label{sec:Introduction}

\IEEEPARstart{I}{n} many signal processing, resource allocation, and machine learning problems, system parameters and loss functions vary over time under dynamic environments. Online learning has emerged as a promising solution to these problems in the presence of uncertainty, where an online decision strategy iteratively adapts to system variations based on historical information~\cite{OL}. Online convex optimization (OCO) is a subclass of online learning, where the loss and constraint functions are convex with respect to (w.r.t.) the decision \cite{OLBK}. OCO can be seen as a sequential decision-making process between a decision maker and the system. Under the standard OCO setting, at the beginning of each time slot, the decision maker selects a decision from a convex feasible set. Only at the end of each time slot, the system reveals  information about
the current convex loss function to the decision maker. The goal of the decision
maker is to minimize the cumulative loss. Such an OCO framework has many
applications, \eg wireless transmit covariance matrix design \cite{UW}, dynamic
network resource allocation \cite{DC}, and smart grids with renewable energy
supply~\cite{SG1}.

In OCO, due to the lack of in-time information about the current convex loss
function, the decision maker cannot select an optimal decision at each time
slot. Instead, the decision maker aims at minimizing the \textit{regret}
\cite{Zinkevich}, \ie the performance gap between the online decision sequence
and some performance benchmark. Most of the early OCO algorithms were evaluated
in terms of the \textit{static} regret, which compares the online decision
sequence with a static offline benchmark that has apriori information of
all the convex loss functions. However, when the environment changes drastically,
the static offline benchmark itself may perform poorly. In this case, the static regret may not be a meaningful performance measurement anymore. A more useful \textit{dynamic} regret measures the performance gap between the online decision sequence and a time-varying sequence of per-time-slot optimizers given knowledge of the current convex loss function. The dynamic regret has been recognized as a more attractive but harder-to-track performance measurement for OCO.

In many practical systems, the decision maker often collects the system parameters
and makes decisions in a \textit{periodic} manner, \eg to limit the computation and communication overhead. One application of interest is precoding design in massive multiple-input multiple-output (MIMO) systems, where the precoder is updated based on delayed channel state information (CSI) feedback and is fixed for a period, \ie one or multiple resource block durations, while the underlying channel can fluctuate quickly over time. The resource block duration is fixed in Long-Term Evolution (LTE) and is allowed to change
over time for a more flexible network operation in 5G New Radio (NR)~\cite{NR}.
In mobile edge computing~\cite{MEC}, due to the offloading and scheduling
latency, the cloud server may periodically collect the offloading tasks from
the remote devices and design a fixed computing resource allocation strategy
for a certain time period.

However, to the best of our knowledge, all existing works on OCO require both the decision and feedback information are updated at each time slot. Motivated by this discrepancy, in this work,
we consider a new constrained OCO problem with \textit{periodic updates},
where the decision maker periodically collects information feedbacks and
makes online decisions to minimize the accumulated loss. The duration of
update period can be multiple time slots and can vary over time. In the
presence of periodic updates, no existing work provides regret bound analysis for OCO.

Furthermore, we consider both short-term and long-term constraints, which are important in many practical optimization problems. For example, in communication systems, the short-term constraint can represent the maximum transmit power, while the long-term power constraint can be seen as a limit on energy usage. An effective constrained OCO algorithm should also bound the \textit{constraint
violation}, which is the accumulated violation on the long-term constraints.
The need to provide the constraint violation bound further adds to the challenges
of regret bound analysis.

The main contributions of this paper are as follows:

\begin{itemize}

\item We formulate a new constrained OCO problem with periodic updates. Each update period may last for multiple time slots
and may vary over time. At the beginning of each update period, the decision maker selects a decision, fixed for the period, to minimize
the accumulated loss subject to both short-term and long-term constraints.
The feedback information about the loss functions can be delayed for multiple
time slots and partly missing.  As explained above, this constrained OCO
framework with periodic updates has broad applications in practical communication
and computation systems.

\item We propose an efficient algorithm, termed Periodic Queueing and Gradient
Aggregation (PQGA) for the formulated constrained OCO problem. In PQGA, we
propose a novel construction of \textit{periodic queues}, which converts
the accumulated constraint violation in an update period into queue dynamics.
Furthermore, PQGA collects and aggregates the delayed gradient feedbacks
in each update period. The periodic queues, together with gradient aggregation,
improve the efficacy of periodic decision updates and facilitate the performance
bounding of PQGA.

\item We analyze the performance of PQGA and study the impact of the periodic queues and gradient aggregation. We prove that PQGA yields $\mathcal{O}(\max\{T^\frac{1+\nu}{2},T^{\delta+\kappa}\})$
dynamic regret, $\mathcal{O}(\max\{T^\frac{1}{2},T^{\delta+\kappa}\})$ static
regret, and $\mathcal{O}(T^{\frac{1}{2}-\kappa})$ constraint violation, where $T$ is the total time horizon, $\nu$ represents the growth rate of the accumulated variation of the per-time-slot optimizer, $\delta$ measures the level of variation of the update period, and $\kappa\in[0,\frac{1}{2}]$ is a tunable trade-off parameter. We further show that, when the number of gradient descent steps within each update period is large enough, PQGA provides improved $\mathcal{O}(\max\{T^\nu,T^\delta\})$ dynamic regret and $\mathcal{O}(1)$ constraint violation. For the special case of per-time-slot updates, PQGA achieves $\mathcal{O}(T^\nu)$ dynamic regret and $\mathcal{O}(1)$ constraint violation bound.

\item As an application, we use PQGA to solve an online precoding design problem in massive MIMO systems with multiple wireless service providers, where all the antennas and wireless spectrum resources are simultaneously shared by the service providers. In this case, we show that PQGA only involves low-complexity closed-form computation. Simulation results show that PQGA converges fast and substantially outperforms the known best alternative.

\end{itemize}

\textit{Organizations:} The rest of this paper is organized as follows. In
Section \ref{Sec:Related Work}, we present the related work. Section \ref{Sec:Problem}
describes the mathematical model, problem formulation, and performance measurement
for constrained OCO with periodic updates. We present PQGA, derive its performance
bounds, and discuss its performance merits in Section \ref{Sec:Algorithm}.
The application of PQGA to large-scale multi-antenna systems with multiple
wireless service providers is presented in Section \ref{Sec:WNV}. Simulation
results are provided in Section \ref{Simulation}, followed by concluding
remarks in Section \ref{Sec:Conclusions}.

\textit{Notations}: The transpose, Hermitian transpose, complex conjugate,
trace, Euclidean norm, Frobenius norm, $L_{\infty}$ norm,  and $L_1$ norm
of a matrix $\mathbf{A}$ are denoted by $\mathbf{A}^T$, $\mathbf{A}^H$, $\mathbf{A}^*$,
$\tr\{\mathbf{A}\}$, $\Vert\mathbf{A}\Vert$, $\Vert\mathbf{A}\Vert_F$, $\Vert\mathbf{A}\Vert_\infty$,
and $\Vert\mathbf{A}\Vert_1$, respectively. The notation $\blkdiag\{\mathbf{A}_1,\dots,\mathbf{A}_n\}$
denotes a block diagonal matrix with diagonal elements being matrices $\mathbf{A}_1,\dots,\mathbf{A}_n$,
$\mathbb{E}\{\cdot\}$ denotes expectation, $\Re\{\cdot\}$ denotes the real
part of the enclosed parameter, $\mathbf{I}$ denotes an identity matrix,
and $\mathbf{g}\sim\mathcal{CN}(\mathbf{0},\sigma^2\mathbf{I})$ means that
$\mathbf{g}$ is a circular complex Gaussian random vector with mean $\mathbf{0}$
and variance~$\sigma^2\mathbf{I}$.


\section{Related Work}
\label{Sec:Related Work}

In this section, we survey existing works on OCO. The differences between
the existing literature and our work are summarized in Table \ref{TAB:comp}.

\subsection{Online Learning and OCO}

Online learning is a method of machine learning, where a learner attempts
to tackle some decision-making task by learning from a sequence of data instances.
As an important subclass of online learning, OCO has been applied in various
areas such as wireless communications~\cite{UW}, cloud networks \cite{DC},
and smart grids \cite{SG1}. In the seminal work of OCO \cite{Zinkevich},
a simple projected gradient descent algorithm achieved $\mathcal{O}(T^{\frac{1}{2}})$
static regret \cite{Zinkevich}. The static regret  was further improved to $\mathcal{O}(\log{T})$ for strongly convex loss functions \cite{OT}. Moreover, \cite{slow} and \cite{AD} examined the static regret for OCO where information feedbacks of the loss functions are delayed for multiple time slots.

The analysis of static regret was extended to that of the more attractive
dynamic regret in \cite{Zinkevich}, \cite{E.C.Hall}, \cite{CDC} for general
convex loss functions. Moreover, strongly convexity was shown to improve
the dynamic regret bound in \cite{Mokhtari16}. By increasing the number of
gradient descent steps, the dynamic regret bound was further improved in
\cite{L.Zhang17}. Furthermore, \cite{Dixit19} studied the impact of inexact
gradient on the dynamic regret bound. \phantom{\cite{Trade}\nocite{LTC-Toff}\nocite{X.CaoTau}-\cite{T.Chen}, \cite{LTC-HY}\nocite{Yu-SC}\nocite{X.Cao}-\cite{INFOCOM21}}

\begin{table*}[t!]
\vspace{-1mm}
\renewcommand{\arraystretch}{1.2}
\caption{Summary of Related Works on OCO}
\vspace{-2mm}
\label{TAB:comp}
\centering
\small
\begin{tabular}{|c|c|c|c|c|c|c|c|c|c|c|c|c|c|c|c|c|c|c|c|}\hline
Reference&Type of benchmark&Long-term constraint&Periodic updates\\\hline\hline
\cite{Zinkevich} &Static and dynamic&No&No\\\hline
\cite{OT}\nocite{slow}-\cite{AD} &Static&No&No\\\hline
\cite{E.C.Hall}\nocite{CDC}\nocite{Mokhtari16}\nocite{L.Zhang17}-\cite{Dixit19}&Dynamic&No&No\\\hline
\cite{Trade}\nocite{LTC-Toff}-\cite{X.CaoTau}, \cite{LTC-HY}, \cite{Yu-SC}&Static&Yes&No\\\hline
\cite{T.Chen}, \cite{X.Cao}&Dynamic&Yes&No\\\hline
\cite{INFOCOM21}&Static and dynamic&Yes&No\\\hline
PQGA &Static and dynamic&Yes&Yes\\\hline
\end{tabular}
\normalsize
\vspace{-4mm}
\end{table*}

\subsection{OCO with Long-Term Constraints}

The above OCO works \cite{Zinkevich}, \cite{OT}\nocite{slow}\nocite{AD}\nocite{E.C.Hall}\nocite{CDC}\nocite{Mokhtari16}\nocite{L.Zhang17}-\cite{Dixit19}
focused on online problems with short-term constraints represented by a feasible
set that must be strictly satisfied. However, long-term constraints arise
in many practical applications such as energy control in wireless communications,
queueing stability in cloud networks, and power balancing in smart grids.
Existing algorithms for OCO with long-term constraints can be categorized
into saddle-point-typed algorithms \cite{Trade}\nocite{LTC-Toff}\nocite{X.CaoTau}-\cite{T.Chen}
and virtual-queue-based algorithms \cite{LTC-HY}\nocite{Yu-SC}\nocite{X.Cao}-\cite{INFOCOM21}.

A saddle-point-typed algorithm was first proposed in \cite{Trade} and achieved
$\mathcal{O}(T^{\frac{1}{2}})$ static regret and $\mathcal{O}(T^{\frac{3}{4}})$
constraint violation. A follow-up work \cite{LTC-Toff} provided $\mathcal{O}(T^{\max\{\mu,1-\mu\}})$
static regret and $\mathcal{O}(T^{1-\frac{\mu}{2}})$ constraint violation,
where $\mu\in(0,1)$ is some trade-off parameter. This recovers the performance
bounds in \cite{Trade} as a special case. In the presence of multi-slot delay,
\cite{X.CaoTau} achieved $\mathcal{O}(T^{\frac{1}{2}})$ static regret and
$\mathcal{O}(T^{\frac{3}{4}})$ constraint violation. The saddle-point-typed
algorithm was further modified in \cite{T.Chen} with dynamic regret analysis.

As an alternative to saddle-point-typed algorithms, virtual queues can be
used to represent the backlog of constraint violation, which facilitates
performance bounding through the analysis of a drift-plus-penalty (DPP) like
expression. A virtual-queue-based algorithm was first proposed in \cite{LTC-HY}
and established $\mathcal{O}(T^{\frac{1}{2}})$ static regret and $\mathcal{O}(1)$
constraint violation for OCO with fixed long-term constraints. For stochastic
constraints that are independent and identically distributed (i.i.d.), another
virtual-queue-based algorithm in \cite{Yu-SC} achieved $\mathcal{O}(T^{\frac{1}{2}})$
static regret and $\mathcal{O}(T^{\frac{1}{2}})$ constraint violation simultaneously.
In \cite{X.Cao}, the virtual-queue-based algorithm was further extended to
provide a dynamic regret bound. The impact of multi-slot feedback delay on
constrained OCO was considered in~\cite{INFOCOM21} with both dynamic and static regret analyses.
 
However, all of the above works on constrained OCO \cite{Trade}\nocite{LTC-Toff}\nocite{X.CaoTau}\nocite{T.Chen}\nocite{LTC-HY}\nocite{Yu-SC}\nocite{X.Cao}-\cite{INFOCOM21}
are under the standard per-time-slot update setting. No other known work
considers periodic updates for OCO. Furthermore, these works only performs single-step gradient descent at each time slot, which does not take full advantage of the potential computational capacity to improve the system performance. In this paper, we propose PQGA, which
uses novel periodic queues with possibly multi-step aggregated gradient descent
to update the online decision. We believe this is the first of its kind.

A part of this work has appeared as a short paper that focuses only on the
application of constrained OCO with period updates to large-scale multi-antenna
systems \cite{SPAWC20}. In the current manuscript, we have substantially
extended our prior work, generalizing the PQGA algorithm, accommodating multi-step gradient descent, deriving new
regret and constraint violation bounds over time-varying update periods,
and providing other new derivations, proofs, and simulation results.

\subsection{Lyapunov Optimization}

PQGA is substantially different from  the conventional DPP algorithm for
Lyapunov optimization~\cite{Neely} in both the decision update and the  virtual
queue update. Lyapunov optimization makes use of the system state and queueing
information to implicitly learn and adapt to system variation with unknown
statistics. The standard Lyapunov optimization is confined to per-time-slot
updates \cite{Neely}. It was extended in \cite{Renewal} to accommodate variable
renewal frames. However, under this framework, the system states are commonly
assumed to be i.i.d. or Markovian, while the OCO framework does not have
such restriction. Furthermore, \cite{Renewal} assumes the system state to
be fixed within each renewal frame, while we allow the loss function to change
at each time slot within an update period.

In addition, the standard Lyapunov optimization relies on the current and
accurate system state for decision updates  \cite{Neely}. When the system
state feedback is delayed, one can apply Lyapunov optimization by leveraging
the historical information to predict the current system state with some
error \cite{Lotfinezhad10}. However, this way of dealing with feedback delay
is equivalent to extending Lyapunov optimization to inaccurate system states
\cite{H.Yu}, \cite{Wang20}. In this case, the optimality gap would be $\mathcal{O}(\sigma{T})$,
where $\sigma$ is some inaccuracy measure.


\section{Constrained OCO with Periodic Updates}
\label{Sec:Problem}

In this section, we detail the mathematical model of constrained OCO with
periodic updates, and we present the formulation of the static regret, dynamic regret, and constraint violation for performance measurement.

\subsection{OCO Problem Formulation}
\label{Sec:Constrained-OCO}

We consider a time-slotted system with time indexed by~$t$. Let $f_t(\mathbf{x}):\mathbb{R}^n\to\mathbb{R}$
be a loss function at time slot $t\in\mathcal{T}=\{0,\dots,T-1\}$. The loss
function $f_t(\mathbf{x})$ is convex and can change arbitrarily over time.
Let $\mathbf{x}_t\in\mathbb{R}^n$ be the decision vector at time slot $t$.
Let $\mathcal{X}_0\subseteq\mathbb{R}^n$ be a compact convex set  that represents
the short-term constraints for any $\mathbf{x}_t$, $t\in\mathcal{T}$. The
goal of the decision maker is to minimize the accumulated loss $\sum_{t\in\mathcal{T}}f_t(\mathbf{x}_t)$.

\begin{figure}[!t]
\centering
\vspace{-0mm}
\includegraphics[width=.7\linewidth,trim=100 320 120 290,clip]{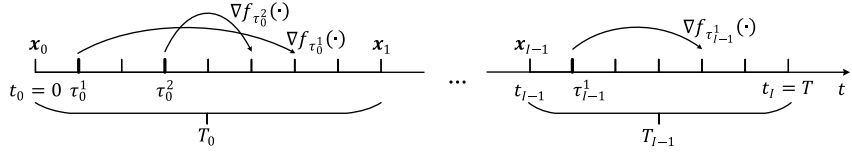}
\vspace{-4mm}
\caption {A timeline illustrating OCO with periodic updates.}
\label{FIG:Periodic}
\vspace{-4mm}
\end{figure}

In standard OCO, the decision maker can update $\mathbf{x}_t$ for any $t\in\mathcal{T}$.
As explained above, this often is not possible in many practical systems. Therefore, in this paper, we consider periodic decision updates for OCO. Suppose the time horizon of $T$ time slots is segmented into $I$ update periods, as shown in Fig.~\ref{FIG:Periodic}. Each update period $i\in\mathcal{I}=\{0,\dots,I-1\}$
has a duration of ${T}_i\in\{1,\dots,T_{\text{max}}\}$ time slots with $T_{\text{max}}$
being the maximum duration of an update period. We have $T=\sum_{i\in\mathcal{I}}T_i$.
Let $t_i$ represent the beginning time slot of update period $i$. The decision
vector is updated at the beginning of time slot $t_i$. For convenience of
exposition, we slightly abuse the notation and use $\mathbf{x}_i$ to denote
this decision vector. It remains unchanged within update period $i$, \ie
$\mathbf{x}_t=\mathbf{x}_i$ for any $t\in\mathcal{T}_i=\{t_i,t_i+1,\dots,t_{i}+T_i-1\}$,
$i\in\mathcal{I}$. Under this new per-period update setting, the accumulated
loss becomes $\sum_{i\in\mathcal{I}}\sum_{t\in\mathcal{T}_i}f_t(\mathbf{x}_i)$.

Let $\nabla{f}_t(\cdot)$, $t\in\mathcal{T}_i$, be the possible gradient information
within update period $i$.  Assume that there are ${S}_i\in\{1,\dots,{T}_i\}$
gradient feedbacks received by the decision maker within update period~$i$.
Let $\tau_i^s$, for $s\in\mathcal{S}_i=\{1,\dots,S_i\}$, represent the time
slot at which the $s$-th gradient feedback in update period $i$, denoted
by $\nabla{f}_{\tau_i^s}(\cdot)$, is sent. The decision maker receives $\nabla{f}_{\tau_i^s}(\cdot)$
after some delay that can last for multiple time slots. Any feedback received
after the next decision $\mathbf{x}_{i+1}$ is assumed to be dropped. Due
to random delays, the gradient feedbacks may be received out of order.

Besides $\mathcal{X}_0$, we also consider long-term constraints on $\{\mathbf{x}_t\}$,
which arise in many practical applications as explained in Section \ref{sec:Introduction}.
Let $\mathbf{g}(\mathbf{x})=[g^1(\mathbf{x}),\dots,g^C(\mathbf{x})]^T:\mathbb{R}^n\to\mathbb{R}^C$
be a vector of $C$ constraint functions. The decision sequence  is subject
to long-term constraints $\sum_{t\in\mathcal{T}}\mathbf{g}(\mathbf{x}_t)\preceq\mathbf{0}$.
With periodic decision updates, it is equivalent to satisfying $\sum_{i\in\mathcal{I}}T_i\mathbf{g}(\mathbf{x}_i)\preceq\mathbf{0}$.

Thus, the goal of constrained OCO with periodic updates is to select a sequence
of decisions $\{\mathbf{x}_i\}$, for $\mathbf{x}_i\in\mathcal{X}_0$, to minimize
the accumulated loss functions while meeting the long-term constraints. This
leads to the following optimization problem:
\begin{align}
        \textbf{P1}:\quad\min_{\{\mathbf{x}_i\}}\quad&\sum_{i\in\mathcal{I}}\sum_{t\in\mathcal{T}_i}f_t(\mathbf{x}_i)\notag\\
        \text{s.t.}~\quad&\sum_{i\in\mathcal{I}}T_i\mathbf{g}(\mathbf{x}_i)\preceq\mathbf{0},\label{EQ:ltc}\\
        &\mathbf{x}_i\in\mathcal{X}_0,\quad\forall{i}\in\mathcal{I}.
\end{align}

Different from existing works on OCO with only short-term constraints \cite{Zinkevich},
\cite{OT}\nocite{slow}\nocite{AD}\nocite{E.C.Hall}\nocite{CDC}\nocite{Mokhtari16}\nocite{L.Zhang17}-\cite{Dixit19},
the additional long-term constraints in (\ref{EQ:ltc}) of \textbf{P1}  lead
to a more complicated online optimization problem, especially since the underlying
system varies over time while the online decision is fixed for a period.
Note that in the special case when update period $T_i=1$ for any $i\in\mathcal{I}$,
\textbf{P1} is simplified to the standard constrained OCO problem with per-time-slot
updates as in \cite{Trade}, \cite{LTC-Toff}, \cite{LTC-HY}.

\subsection{Performance Metric}
\label{Sec:Performance Metrics}

Due to the lack of in-time information about the current loss functions under
the OCO setting, an optimal solution to \textbf{P1} cannot be obtained.\footnote{In
fact, even for the simplest original OCO problem \cite{Zinkevich} (\ie  under
the per-time-slot update setting without long-term constraints (\ref{EQ:ltc})),
an optimal solution cannot be found \cite{OT}.} We consider the following
performance measurements typically adopted for developing the solution for
constrained OCO, with a slight modification tailored to periodic updates.

We aim at designing a decision sequence $\{\mathbf{x}_i\}$ over update periods,
such that the accumulated loss in the objective of \textbf{P1} is competitive
with some benchmark under the same set of gradient feedbacks. Thus, for the static regret, we consider the following static offline benchmark, which is generalized from the per-time-slot one used in \cite{Trade}\nocite{LTC-Toff}-\cite{X.CaoTau}, \cite{LTC-HY}, \cite{Yu-SC} to accommodate periodic updates:
\begin{align}
        \mathbf{x}^\star\triangleq\arg\min_{\mathbf{x}\in\mathcal{X}}\sum_{i\in\mathcal{I}}\frac{T_i}{S_i}\sum_{s\in\mathcal{S}_i}f_{\tau_i^s}(\mathbf{x})\label{EQ:xopt}
\end{align}
where $\mathcal{X}\triangleq\{\mathbf{x}\in\mathcal{X}_0:\mathbf{g}(\mathbf{x})\preceq\mathbf{0}\}$.
 Note that $\mathbf{x}^\star$ is computed offline assuming all the loss functions
$f_{\tau_i^s}(\mathbf{x})$, for all $s\in\mathcal{S}_i$ and $i\in\mathcal{I}$,
are known in advance. Then, the static regret is the performance gap between
$\{\mathbf{x}_i\}$ and $\mathbf{x}^\star$:
\begin{align}
        \text{RE}_{\text{s}}(T)\triangleq\sum_{i\in\mathcal{I}}\frac{T_i}{S_i}\sum_{s\in\mathcal{S}_i}\left(f_{\tau_i^s}(\mathbf{x}_i)-f_{\tau_i^s}(\mathbf{x}^\star)\right).\label{EQ:REs}
\end{align}

However, the static regret only provides a coarse performance measure when
the underlying system is time-varying and may not be an attractive metric
to use. A more useful performance benchmark is the dynamic benchmark $\{\mathbf{x}_i^\circ\}$,
given by
\begin{align}
        \mathbf{x}_i^\circ\triangleq\arg\min_{\mathbf{x}\in\mathcal{X}}\frac{T_i}{S_i}\sum_{s\in\mathcal{S}_i}f_{\tau_i^s}(\mathbf{x}).\label{EQ:xtopt}
\end{align}
For the case of per-time-slot updates, the dynamic benchmark was originally
proposed for OCO with short-term constraints \cite{Zinkevich} and was modified
in \cite{T.Chen}, \cite{X.Cao}, \cite{INFOCOM21} to incorporate long-term
constraints. Here, we generalize it to account for periodic updates. In (\ref{EQ:xtopt}),
$\mathbf{x}_i^\circ$ is computed using all the $S_i$ loss functions $f_{\tau_i^s}(\mathbf{x})$
in the current update period $i$. Then, the dynamic regret is
\begin{align}
        \text{RE}_{\text{d}}(T)\triangleq\sum_{i\in\mathcal{I}}\frac{T_i}{S_i}\sum_{s\in\mathcal{S}_i}\left(f_{\tau_i^s}(\mathbf{x}_i)-f_{\tau_i^s}(\mathbf{x}_i^\circ)\right).\label{EQ:RE}
\end{align}
The gap between the static and dynamic regrets can be as large as $\mathcal{O}(T)$
\cite{Gap}. In this paper, for comprehensive performance analysis, we provide
upper bounds for both $\text{RE}_{\text{s}}(T)$ and $\text{RE}_{\text{d}}(T)$.

\begin{remark}
Note that with incomplete gradient feedbacks, \ie $\sum_{i\in\mathcal{I}}S_i=S<T$,
our regret definitions in (\ref{EQ:REs}) and (\ref{EQ:RE}) fully utilize
the feedback information. Our accumulated loss in each period $i$ is the
average loss over time slots when the gradient feedbacks are provided, \ie $\frac{1}{S_i}\sum_{s\in\mathcal{S}_i}f_{\tau_i^s}(\mathbf{x}_i)$,
scaled by the duration $T_i$ of the update period $i$, for $i\in\mathcal{I}$.
If the environment is mean stationary, \ie $\mathbb{E}\{f_t(\mathbf{x})\}=\mathbb{E}\{f_{t'}(\mathbf{x})\}$
for any $t,t'\in\mathcal{T}$, then in the expectation sense, the accumulated
loss in our regret definitions is the same as the actual loss in  the objective
of \textbf{P1}, \ie
\begin{align*}
        \mathbb{E}\left\{\sum_{i\in\mathcal{I}}\frac{T_i}{S_i}\sum_{s\in\mathcal{S}_i}\left(f_{\tau_i^s}(\mathbf{x}_i)\right)\right\}=\mathbb{E}\left\{\sum_{i\in\mathcal{I}}\sum_{t\in\mathcal{T}_i}\left(f_t(\mathbf{x}_i)\right)\right\}.
\end{align*}
More generally, suppose there exists a constant $d>0$ such that $|f_t(\mathbf{x})-f_{t'}(\mathbf{x})|\le{d}$,
for any $\mathbf{x}\in\mathcal{X}_0$ and $t,t'\in\mathcal{T}_i$, $i\in\mathcal{I}$.
Then, we have $\sum_{i\in\mathcal{I}}\frac{T_i}{S_i}\sum_{s\in\mathcal{S}_i}(f_{\tau_i^s}(\mathbf{x}_i))-\sum_{i\in\mathcal{I}}\sum_{t\in\mathcal{T}_i}(f_t(\mathbf{x}_i))=\mathcal{O}(d(T-S))$.
This performance gap can be small if the system does not fluctuate too much
over time. Furthermore, it approaches zero as the number of feedbacks $S\to{T}$.
\end{remark}

We also need to measure the accumulated violation of each long-term constraint
$c\in\mathcal{C}=\{1,\dots,C\}$. Define the constraint violation as
\begin{align}
        \text{VO}^c(T)\triangleq\sum_{i\in\mathcal{I}}T_ig^c(\mathbf{x}_i),\quad\forall{c}\in\mathcal{C}.\label{EQ:VO}
\end{align}
Note that the constraint violation defined in \cite{Trade}\nocite{LTC-Toff}\nocite{X.CaoTau}\nocite{T.Chen}\nocite{LTC-HY}\nocite{Yu-SC}\nocite{X.Cao}-\cite{INFOCOM21}
are under the standard per-time-slot update setting. In contrast, our model
accommodates the possibly time-varying update periods of multiple time slots.

\section{The Periodic Queueing and Gradient Aggregation (PQGA) Algorithm}
\label{Sec:Algorithm}

In this section, we present an efficient algorithm, PQGA, to solve the formulated
constrained OCO problem. It uses a periodic virtual queue and periodic updates after solving per-period optimization problems that are convex and hence practically solvable. We further show that, despite its low implementation complexity, PQGA provides provable performance guarantees in terms of dynamic regret, static regret, and constraint violation
bounds.
\subsection{PQGA Algorithm Description}

PQGA introduces a \textit{periodic} virtual queue vector $\mathbf{Q}_i=[Q_i^1,\dots,Q_i^C]^T$
in each update period $i\in\mathcal{I}$, with the following updating rule:
\begin{align}
        Q_{i+1}^c=\max\{-\gamma{T}_i{g}^c(\mathbf{x}_i),Q_i^c+\gamma{T}_i{g}^c(\mathbf{x}_i)\},\quad\forall{c}\in\mathcal{C}\label{EQ:VQ}
\end{align}
where $\gamma>0$ is an algorithm parameter. The role of $\mathbf{Q}_i$ is
similar to a Lagrange multiplier vector for the long-term constraints in
(\ref{EQ:ltc}), and the value of $\mathbf{Q}_i$ reflects the accumulated
violation of the long-term constraints. We remark here that (\ref{EQ:VQ})
is different from the virtual queues used in the standard Lyapunov optimization
\cite{Neely} and subsequent extensions to OCO \cite{LTC-HY}\nocite{Yu-SC}\nocite{X.Cao}-\cite{INFOCOM21}.
Unique to our proposed approach, ${T}_i{g}^c(\mathbf{x}_i)$ is the accumulated
constraint violation in update period $i$, and it is scaled by an appropriate
factor $\gamma$.

In the basic form of PQGA, instead of solving \textbf{P1} directly, we solve a \textit{per-period} problem
at the beginning of each update period $i+1$ with the short-term constraints
only, given by
\begin{align*}
        \textbf{P2}:~\min_{\mathbf{x}\in\mathcal{X}_0}~\frac{T_i}{S_i}\sum_{s\in\mathcal{S}_i}[\nabla{f}_{\tau_i^s}(\mathbf{x}_i)]^T(\mathbf{x}-\mathbf{x}_i)+\alpha\Vert\mathbf{x}-\mathbf{x}_i\Vert^2+[\mathbf{Q}_{i+1}+\gamma{T}_i\mathbf{g}(\mathbf{x}_i)]^T[\gamma{T}_{i+1}\mathbf{g}(\mathbf{x})]
\end{align*}
where $\alpha,\gamma>0$ are two algorithm parameters. In \textbf{P2}, the
gradient direction is aggregated based on all the gradient feedbacks $\{\nabla{f}_{\tau_i^s}(\cdot),s\in\mathcal{S}_i\}$,
collected in the previous update period~$i$. The regularization term $\alpha\Vert\mathbf{x}-\mathbf{x}_i\Vert^2$
controls how much the new decision $\mathbf{x}_{i+1}$ is allowed to change
 from the previous decision $\mathbf{x}_{i}$. Furthermore, in the last term of the objective
function of \textbf{P2}, we consider an inner-product between the vector
of periodic queue lengths and the period-weighted vector of the long-term
constraint functions, which represents a penalty of constraint violation
in $\mathbf{g}(\mathbf{x})$. Thus, we convert the long-term constraints in
(\ref{EQ:ltc}) to a penalty term on $\mathbf{g}(\mathbf{x})$ as one part
of the objective function in \textbf{P2}.

In addition to the basic form of PQGA that uses a single step of gradient descent in \textbf{P2}, we can further configure PQGA to incorporate multi-step gradient descent. Multi-step gradient descent has previously been shown to provide stronger optimization results for OCO with short-term constraints \cite{L.Zhang17}. In this work, we will further verify that it also provides performance improvement to PQGA under both short-term and long-term constraints. Specifically, at the beginning of each update period $i+1$, after updating the periodic virtual queue in (\ref{EQ:VQ}), we initialize an intermediate decision vector $\tilde{\mathbf{x}}_{i}^0=\mathbf{x}_i$. We then perform $J$-step aggregated gradient descent to generate $\tilde{\mathbf{x}}_{i}^J$ for any $J\ge0$. If $J=0$, we readily have $\tilde{\mathbf{x}}_i^J=\mathbf{x}_i$ by initialization. Otherwise, for each gradient descent step $j\in\mathcal{J}=\{1,\dots,J\}$, we solve the following optimization problem for $\tilde{\mathbf{x}}_{i}^j$:
\begin{align*}
        \min_{\mathbf{x}\in\mathcal{X}_0}~\frac{T_i}{S_i}\sum_{s\in\mathcal{S}_i}[\nabla{f}_{\tau_i^s}(\tilde{\mathbf{x}}_{i}^{j-1})]^T(\mathbf{x}-\tilde{\mathbf{x}}_{i}^{j-1})+\alpha\Vert\mathbf{x}-\tilde{\mathbf{x}}_{i}^{j-1}\Vert^2.
\end{align*}
The above problem is similar to the standard projected gradient descent problem. Therefore, its solution is readily available:
\begin{align}
        \tilde{\mathbf{x}}_{i}^j=\mathcal{P}_{\mathcal{X}_0}\left\{\tilde{\mathbf{x}}_{i}^{j-1}-\frac{1}{2\alpha}\left(\frac{T_i}{{S}_i}\sum_{s\in\mathcal{S}_i}\nabla{f}_{\tau_i^s}(\tilde{\mathbf{x}}_{i}^{j-1})\right)\right\}\label{eq:xij}
\end{align}
where $\mathcal{P}_{\mathcal{X}_0}\{\mathbf{x}\}\triangleq\arg\min_{\mathbf{y}\in\mathcal{X}_0}\{\Vert\mathbf{y}-\mathbf{x}\Vert^2\}$
is the projection operator onto the convex feasible set $\mathcal{X}_0$ and
$\alpha$ can be viewed as a step-size parameter.

With both $\mathbf{x}_i$ and $\tilde{\mathbf{x}}_i^J$, we then modify
$\textbf{P2}$ to the following per-period optimization problem for
$\mathbf{x}_{i+1}$:
\begin{align*}
        \textbf{P2}':\min_{\mathbf{x}\in\mathcal{X}_0}\frac{T_i}{S_i}\sum_{s\in\mathcal{S}_i}[\nabla{f}_{\tau_i^s}(\tilde{\mathbf{x}}_i^J)]^T(\mathbf{x}\!-\!\tilde{\mathbf{x}}_i^J)\!+\!\alpha\Vert\mathbf{x}\!-\!\tilde{\mathbf{x}}_i^{J}\Vert^2\!+\!\eta\Vert\mathbf{x}\!-\!\mathbf{x}_{i}\Vert^2\!+\![\mathbf{Q}_{i+1}\!+\!\gamma{T}_i\mathbf{g}(\mathbf{x}_i)]^T[\gamma{T}_{i+1}\mathbf{g}(\mathbf{x})]
\end{align*}
where $\alpha,\eta,\gamma>0$ and $J\ge0$ are four algorithm parameters. Note that $\textbf{P2}'$ uses double regularization on both $\mathbf{x}_i$ and  $\tilde{\mathbf{x}}_i^J$. The intuition behind the double regularization is that both $\mathbf{x}_i$ and  $\tilde{\mathbf{x}}_i^J$ help to minimize the accumulate loss and constraint violation. Therefore, it is beneficial to prevent the new decision $\mathbf{x}_{i+1}$ from being too far away from either of them. This will be shown, analytically in Section \ref{Sec:Bounds}, to give PQGA substantial performance advantage over existing works in terms of performance bounds.

The PQGA algorithm is given in Algorithm \ref{Alg:PQGA}. Note that PQGA has
four algorithm parameters $\alpha,\eta,\gamma$, and $J$. We will  discuss the choice of their values in Section \ref{Sec:Discuss}, after we derive the regret and constraint violation bounds, to explain the impact of these four parameters on those bounds.  During each update period $i\in\mathcal{I}$, the decision maker collects the delayed and possibly incomplete gradient information $\nabla{f}_{\tau_i^s}(\cdot),s\in\mathcal{S}_i$. At the beginning of the next update period $i+1$, it first updates the periodic virtual queue $\mathbf{Q}_{i+1}$ in (\ref{EQ:VQ}) based on the accumulated constraint violation caused by its previous decision $\mathbf{x}_i$. It then learns the gradient descent direction from the collected past gradient information and performs $J$-step aggregated gradient descent to generate $\tilde{\mathbf{x}}^J$. Finally, it weights the constraint functions based on the updated queue lengths
and regularizes on both $\tilde{\mathbf{x}}^J$ and $\mathbf{x}_i$, to compute
the decision $\mathbf{x}_{i+1}$ for update period $i+1$ by solving $\textbf{P2}'$.\footnote{When
$J=0$, we readily have $\tilde{\mathbf{x}}^J=\mathbf{x}_i$ by initialization.
In this case, the double regularization in $\textbf{P2}'$ on $\tilde{\mathbf{x}}^J$
and $\mathbf{x}_i$ can be combined as a single regularization on $\mathbf{x}_i$,
and $\textbf{P2}'$ is equivalent to $\textbf{P2}$.}

\begin{remark}
$\textbf{P2}$ and $\textbf{P2}'$ are strongly convex optimization problems, so they can be solved efficiently using well-known optimization tools. Furthermore, as shown in Section \ref{Sec:Online Precoding Solution}, for the considered problem of  precoding-based massive MIMO virtualization, it has a closed-form solution with negligible computational complexity.
\end{remark}

\begin{remark}
When the vector long-term constraint function $\mathbf{g}(\mathbf{x})$ is
separable w.r.t. $\mathbf{x}$, $\textbf{P2}$ and $\textbf{P2}'$ can be equivalently decomposed
into independent subproblems. In this case, PQGA can be implemented distributively
with even lower computational complexity.
\end{remark}

\begin{algorithm}[!t]
\caption{The PQGA Algorithm}
\label{Alg:PQGA}
\begin{algorithmic}[1]
\STATE {\textbf{Initialization}: $\alpha,\eta,\gamma>0$, $J\ge0$, $\mathbf{x}_0\in\mathcal{X}_0$,
and $\mathbf{Q}_0=\mathbf{0}$.}
\STATE {At the beginning of each update period $i+1$, do:}
\STATE {\quad Update the periodic virtual queue $\mathbf{Q}_{i+1}$ via (\ref{EQ:VQ}).}
\STATE {\quad Initialize the intermediate decision vector $\tilde{\mathbf{x}}_{i}^0=\mathbf{x}_i$.}
\STATE {\quad\textbf{for} $j=1$ \textbf{to} $J$}
\STATE {\qquad Update $\tilde{\mathbf{x}}^j$ via (\ref{eq:xij}).}
\STATE {\quad\textbf{end for}}
\STATE {\quad Update the periodic decision $\mathbf{x}_{i+1}$ by solving~$\textbf{P2}'$\\
        \quad using $\mathbf{Q}_{i+1}$, $\mathbf{x}_i$, and $\tilde{\mathbf{x}}^J$.}
\end{algorithmic}
\end{algorithm}

\subsection{Regret and Constraint Violation Bounds of PQGA}
\label{Sec:Bounds}

Existing analysis techniques for the standard per-time-slot OCO setting with single-step gradient descent \cite{Trade}\nocite{LTC-Toff}\nocite{X.CaoTau}\nocite{T.Chen}\nocite{LTC-HY}\nocite{Yu-SC}\nocite{X.Cao}-\cite{INFOCOM21} are inadequate for studying the performance of PQGA. In this section, we
present new techniques to derive the regret and constraint violation
bounds of PQGA, particularly to account for the periodic queues and possibly multi-step aggregated gradient descent. Although a small part of our derivations uses techniques from
Lyapunov drift analysis, as explained in Section \ref{Sec:Related Work},
PQGA is structurally different from Lyapunov optimization.

We make the following assumptions that are common in the literature of OCO.

\begin{assumption}\label{ASP-strong}
For any $t$, the loss function $f_t(\mathbf{x})$ satisfies the following:
\begin{enumerate}
\item[1.1)] $f_t(\mathbf{x})$ is $2\varrho$-strongly convex over $\mathcal{X}_0$:
$\exists~\varrho>0$, s.t., for any $\mathbf{x},\mathbf{y}\in\mathcal{X}_0$ and
$t$
\begin{align}
        \!\!\!\!\!f_t(\mathbf{y})\ge f_t(\mathbf{x})+[\nabla{f}_t(\mathbf{x})]^T(\mathbf{y}-\mathbf{x})+\varrho\Vert\mathbf{y}-\mathbf{x}\Vert^2.\!\!\label{EQ:varrho}
\end{align}
\item[1.2)] $f_t(\mathbf{x})$ is $2L$-smooth over $\mathcal{X}_0$: $\exists~{L}>0$, s.t., for any $\mathbf{x},\mathbf{y}\in\mathcal{X}_0$ and $t$
\begin{align}
        \!\!\!\!\!\!f_t(\mathbf{y})\le f_t(\mathbf{x})+[\nabla{f}_t(\mathbf{x})]^T(\mathbf{y}-\mathbf{x})+L\Vert\mathbf{y}-\mathbf{x}\Vert^2.\!\!\label{EQ:L}
\end{align}
\end{enumerate}
\end{assumption}

\begin{assumption}\label{ASP-1}
The gradient $\nabla{f}_t(\mathbf{x})$ is bounded: $\exists~{D}>0$, s.t.,
\begin{align}
        \Vert\nabla{f}_t(\mathbf{x})\Vert\le D,\quad\forall\mathbf{x}\in\mathcal{X}_0,\quad\forall{t}\in\mathcal{T}.\label{EQ:D}
\end{align}
\end{assumption}

\begin{assumption}\label{ASP-2}
The long-term constraint functions satisfy the following:
\begin{enumerate}
\item[3.1)] $\mathbf{g}(\mathbf{x})$ is Lipschitz continuous on $\mathcal{X}_0$:
$\exists~\beta>0$, s.t.,
\begin{align}
          \Vert\mathbf{g}(\mathbf{x})-\mathbf{g}(\mathbf{y})\Vert\le\beta\Vert\mathbf{x}-\mathbf{y}\Vert,\quad\forall\mathbf{x},\mathbf{y}\in\mathcal{X}_0.\label{EQ:Beta}
\end{align}
\item[3.2)] $\mathbf{g}(\mathbf{x})$ is bounded: $\exists~{G}>0$, s.t.,
\begin{align}
          \Vert\mathbf{g}(\mathbf{x})\Vert\le{G},\quad\forall\mathbf{x}\in\mathcal{X}_0.\label{EQ:G}
\end{align}
\item[3.3)] Existence of an interior point: $\exists~\epsilon>0$ and $\mathbf{x}'\in\mathcal{X}_0$, s.t.,
\begin{align}
        \mathbf{g}(\mathbf{x}')\preceq-\epsilon\mathbf{1}.\label{EQ:Epsilon}
\end{align}
\end{enumerate}
\end{assumption}

\begin{assumption}\label{ASP-3}
The radius of $\mathcal{X}_0$ is bounded: $\exists~{R}>0$, s.t.,
\begin{align}
        \Vert\mathbf{x}-\mathbf{y}\Vert\le R,\quad\forall\mathbf{x},\mathbf{y}\in\mathcal{X}_0.\label{EQ:R}
\end{align}
\end{assumption}

\begin{remark}
Strongly convex loss functions arise in many machine learning and signal
processing applications, such as Lasso regression, support vector machine,
softmax classifier, and robust subspace tracking. Furthermore, for applications
with general convex loss functions, it is common to add a simple regularization
term such as $\mu\Vert\mathbf{x}\Vert^2$, so that the overall optimization
objective becomes strongly convex \cite{Dixit19}.
\end{remark}

\subsubsection{ Bounding the Dynamic Regret}

A main goal of this paper is to examine the impact of possibly time-varying
update periods and multi-step aggregated gradient descent on the dynamic regret bound for constrained OCO, which has not been addressed in the existing
literature. To this end, we define the accumulated variation of the dynamic
benchmark $\{\mathbf{x}_i^\circ\}$ (termed the path length in \cite{Zinkevich})
as
\begin{align}
        \Pi_{\mathbf{x}^\circ}\triangleq\sum_{i\in\mathcal{I}}\Vert\mathbf{x}_i^\circ-\mathbf{x}_{i+1}^\circ\Vert.\label{EQ:PL}
\end{align}
Another related quantity regarding the accumulated variation of the time-varying
update periods $\{T_i\}$ is defined as
\begin{align}
        \Pi_T\triangleq\sum_{i\in\mathcal{I}}(T_i-T_{i+1})^2.\label{EQ:PiT}
\end{align}

We first provide bounds on the periodic virtual queues $\{\mathbf{Q}_i\}$
produced by PQGA in the following lemma.

\begin{lemma}\label{LM-VQ}
The following statements hold for any $i\in\mathcal{I}$:
\begin{align}
        &\mathbf{Q}_i\succeq\mathbf{0},\label{EQ:VQ1}\\
        &\mathbf{Q}_{i+1}+\gamma{T}_i\mathbf{g}(\mathbf{x}_i)\succeq\mathbf{0},\label{EQ:VQ2}\\
        &\Vert\mathbf{Q}_{i+1}\Vert\ge\Vert\gamma{T}_i\mathbf{g}(\mathbf{x}_i)\Vert,\label{EQ:VQ3}\\
        &\Vert\mathbf{Q}_{i+1}\Vert\le\Vert\mathbf{Q}_i\Vert+\Vert\gamma{T}_i\mathbf{g}(\mathbf{x}_i)\Vert.\label{EQ:VQ4}
\end{align}
\end{lemma}
\textit{Proof:}  The periodic virtual queue vector is initialized as $\mathbf{Q}_0=\mathbf{0}$.
For any $c\in\mathcal{C}$ and $i\in\mathcal{I}$, by induction, we first assume
$Q_{i}^c\ge0$. From the periodic virtual queue dynamics in (\ref{EQ:VQ}),
if ${\gamma T}_i{g}^c(\mathbf{x}_i)\ge0$, then $Q_{i+1}^c\ge{Q}_{i}^c+{\gamma{T}}_i{g}^c(\mathbf{x}_i)\ge0$;
otherwise, we have $Q_{i+1}^c\ge-\gamma{T}_i{g}^c(\mathbf{x}_i)\ge0$. Combining
the above two cases, we have (\ref{EQ:VQ1}).

From (\ref{EQ:VQ}), for any $c\in\mathcal{C}$ and $i\in\mathcal{I}$, we have
$Q_{t+1}^c\ge-\gamma{T}_i{g}^c(\mathbf{x}_i)$, which yields (\ref{EQ:VQ2}).

For any $c\in\mathcal{C}$ and $i\in\mathcal{I}$, from (\ref{EQ:VQ}) and $Q_i^c\ge0$
in (\ref{EQ:VQ1}), if ${\gamma T}_i{g}^c(\mathbf{x}_i)\ge0$, then $Q_{i+1}^c\ge{Q}_i^c+{\gamma
T}_i{g}^c(\mathbf{x}_i)\ge{\gamma T}_i{g}^c(\mathbf{x}_i)$; otherwise, we
have $Q_{i+1}^c\ge-\gamma{T}_i{g}^c(\mathbf{x}_i)$. Therefore, we have $Q_{i+1}^c\ge|\gamma{T}_i{g}^c(\mathbf{x}_i)|$.
Squaring both sides and summing over $c\in\mathcal{C}$ yields (\ref{EQ:VQ3}).

From (\ref{EQ:VQ}), for any $c\in\mathcal{C}$ and $i\in\mathcal{I}$, we have
$Q_{i+1}^c\le{Q}_i^c+|\gamma{T}_i{g}^c(\mathbf{x}_i)|$. Since $Q_i^c\ge0$
in (\ref{EQ:VQ1}), by the triangle inequality, we have $\Vert\mathbf{Q}_{i+1}\Vert\le\sqrt{\sum_{c\in\mathcal{C}}(Q_i^c+|\gamma{T}_i{g}^c(\mathbf{x}_i)|)^2}\le\Vert\mathbf{Q}_i\Vert+\Vert\gamma{T}_i\mathbf{g}(\mathbf{x}_i)\Vert$,
which yields (\ref{EQ:VQ4}).
\hfill$\blacksquare$
\vspace*{.6em}

Define $L_i\triangleq\frac{1}{2}\Vert\mathbf{Q}_i\Vert^2$ as the quadratic
Lyapunov function and $\Delta_i\triangleq{L}_{i+1}-L_i$ as the Lyapunov drift
for each update period $i\in\mathcal{I}$. Leveraging the results in Lemma~\ref{LM-VQ},
we provide an upper bound on the Lyapunov drift $\Delta_i$ in the following
lemma. 

\begin{lemma}\label{LM-Drift}
The Lyapunov drift is upper bounded for any $i\in\mathcal{I}$ as follows:
\begin{align}
        \Delta_i\le\mathbf{Q}_i^T[\gamma{T}_i\mathbf{g}(\mathbf{x}_i)]+\Vert\gamma{T}_i\mathbf{g}(\mathbf{x}_i)\Vert^2.\label{EQ:Drift}
\end{align}
\end{lemma}
\textit{Proof:} For any $c\in\mathcal{C}$ and $i\in\mathcal{I}$, we first
prove
\small
\begin{align}
        \frac{1}{2}(Q_{i+1}^c)^2-\frac{1}{2}(Q_i^c)^2\le{Q}_i^c[\gamma{T}_ig^c(\mathbf{x}_i)]+[\gamma{T}_i{g}^c(\mathbf{x}_i)]^2\label{EQ:LM-Drift-1}
\end{align}
\normalsize
by considering the following two cases.

\textit{1) $Q_i^c+{\gamma T}_i{g}^c(\mathbf{x}_i)\ge-\gamma{T}_i{g}^c(\mathbf{x}_i)$:}
From the virtual queue dynamics in (\ref{EQ:VQ}), we have $Q_{i+1}^c=Q_i^c+{\gamma
T}_i{g}^c(\mathbf{x}_i)$.
It then follows that
\small
\begin{align*}
        \frac{1}{2}(Q_{i+1}^c)^2&=\frac{1}{2}[Q_i^c+{\gamma T}_i{g}^c(\mathbf{x}_i)]^2\le\frac{1}{2}(Q_i^c)^2+Q_i^c[\gamma{T}_ig^c(\mathbf{x}_i)]+[\gamma{T}_i{g}^c(\mathbf{x}_i)]^{2}.
\end{align*}
\normalsize

\textit{2) $ Q_i^c+{\gamma T}_i{g}^c(\mathbf{x}_i)<-\gamma{T}_i{g}^c(\mathbf{x}_i)$:}
We have $Q_{i+1}^c=-\gamma{T}_i{g}^c(\mathbf{x}_i)$ from (\ref{EQ:VQ}). It
then follows that
\small
\begin{align*}
        \frac{1}{2}(Q_{i+1}^c)^2&\le\frac{1}{2}[\gamma{T}_i{g}^c(\mathbf{x}_i)]^2+\frac{1}{2}[Q_i^c+\gamma{T}_i{g}^c(\mathbf{x}_i)]^2=\frac{1}{2}(Q_i^c)^2+{Q}_i^c[\gamma{T}_ig^c(\mathbf{x}_i)]+[\gamma{T}_i{g}^c(\mathbf{x}_i)]^{2}.
\end{align*}
\normalsize

Combining the above two cases, we have (\ref{EQ:LM-Drift-1}). Summing (\ref{EQ:LM-Drift-1})
over $c\in\mathcal{C}$ yields (\ref{EQ:Drift}).\hfill$\blacksquare$
\vspace*{.5em}

We also require the following two lemmas, which are reproduced from Lemma 2.8 in \cite{OLBK} and Lemma 1 in \cite{L.Zhang17}, respectively.

\begin{lemma}\label{LM-StronglyConvex}
Let $\mathcal{Z}\subseteq\mathbb{R}^n$ be a nonempty convex set. Let $h(\mathbf{z}):\mathbb{R}^n\to\mathbb{R}$
be a $2\varsigma$-strongly-convex function over $\mathcal{Z}$ w.r.t. $\Vert\cdot\Vert$.
Let $\mathbf{w}=\arg\min_{\mathbf{z}\in\mathcal{Z}}\{h(\mathbf{z})\}$. Then,
for any $\mathbf{u}\in\mathcal{Z}$, we have $h(\mathbf{w})\le{h}(\mathbf{u})-\varsigma\Vert\mathbf{u}-\mathbf{w}\Vert^2$.
\end{lemma}

\begin{lemma}\label{LM-MGD}
Let $\mathcal{Z}\subseteq\mathbb{R}^n$ be a nonempty convex set. Let $h(\mathbf{z}):\mathbb{R}^n\to\mathbb{R}$
be a $2\varsigma$-strongly-convex and $2\zeta$-smooth function over $\mathcal{Z}$
w.r.t. $\Vert\cdot\Vert$. Let $\mathbf{v}=\arg\min_{\mathbf{z}\in\mathcal{Z}}\{[\nabla{h}(\mathbf{u})]^T(\mathbf{z}-\mathbf{u})+\upsilon\Vert\mathbf{z}-\mathbf{u}\Vert^2\}$
and $\mathbf{w}=\arg\min_{\mathbf{z}\in\mathcal{Z}}\{h(\mathbf{z})\}$.
Then, for any $\upsilon\ge\zeta$, we have $\Vert\mathbf{w}-\mathbf{v}\Vert^2\le\frac{\upsilon-\varsigma}{\upsilon+\varsigma}\Vert\mathbf{w}-\mathbf{u}\Vert^2$.
\end{lemma}

Based on Lemmas \ref{LM-VQ}-\ref{LM-MGD}, for any number of aggregated gradient descent steps $J\ge0$, we provide an upper bound on the dynamic regret $\text{RE}_{\text{d}}(T)$ for PQGA in the following theorem.

\begin{theorem}\label{THM-RE}

For any $J\ge0$, if we choose $\alpha\ge{T}_{\text{max}}L$,
$\eta\ge\beta^2\gamma^2T_{\text{max}}^2$, and $\gamma>0$,
the dynamic regret of PQGA is upper bounded by
\begin{align}
        \text{RE}_{\text{d}}(T)&\le\frac{D^2T_{\text{max}}}{4\alpha}T+(\alpha\rho^J+\eta)({R}^2+2{R}\Pi_{\mathbf{x}^\circ})+\gamma^2G^2(T_{\text{max}}^2+\Pi_T)\label{EQ:RE-BD}
\end{align}
where $\rho=\frac{\alpha-\varrho}{\alpha+\varrho}<1$.
\end{theorem}
\textit{Proof:} The objective function of $\textbf{P2}'$ is $2(\alpha+\eta)$-strongly
convex
over $\mathcal{X}_0$ w.r.t. $\Vert\cdot\Vert$ due to the double regularization.
Since
$\mathbf{x}_{i+1}$ minimizes $\textbf{P2}'$ over $\mathcal{X}_0$ for any
$i\in\mathcal{I}$,
we have
\small
\begin{align}
        &\frac{T_i}{S_i}\sum_{s\in\mathcal{S}_i}[\nabla{f}_{\tau_i^s}(\tilde{\mathbf{x}}_i^J)]^T(\mathbf{x}_{i+1}-\tilde{\mathbf{x}}_i^J)+\alpha\Vert\mathbf{x}_{i+1}-\tilde{\mathbf{x}}_i^{J}\Vert^2+[\mathbf{Q}_{i+1}+\gamma{T}_i\mathbf{g}(\mathbf{x}_i)]^T[\gamma{T}_{i+1}\mathbf{g}(\mathbf{x}_{i+1})]+\eta\Vert\mathbf{x}_{i+1}-\mathbf{x}_{i}\Vert^2\notag\\
        &\quad\stackrel{(a)}{\le}\frac{T_i}{S_i}\sum_{s\in\mathcal{S}_i}[\nabla{f}_{\tau_i^s}(\tilde{\mathbf{x}}_i^J)]^T(\mathbf{x}_i^\circ-\tilde{\mathbf{x}}_i^J)+\alpha\Vert\mathbf{x}_i^\circ-\tilde{\mathbf{x}}_i^{J}\Vert^2+[\mathbf{Q}_{i+1}+\gamma{T}_i\mathbf{g}(\mathbf{x}_i)]^T[\gamma{T}_{i+1}\mathbf{g}(\mathbf{x}_i^\circ)]+\eta\Vert\mathbf{x}_i^\circ-\mathbf{x}_{i}\Vert^2\notag\\
        &\qquad-(\alpha+\eta)\Vert\mathbf{x}_i^\circ-\mathbf{x}_{i+1}\Vert^2\notag\\
        &\quad\stackrel{(b)}{\le}\frac{T_i}{S_i}\sum_{s\in\mathcal{S}_i}[\nabla{f}_{\tau_i^s}(\tilde{\mathbf{x}}_i^J)]^T(\mathbf{x}_i^\circ-\tilde{\mathbf{x}}_i^J)+\alpha(\Vert\mathbf{x}_i^\circ-\tilde{\mathbf{x}}_i^{J}\Vert^2-\Vert\mathbf{x}_i^\circ-\mathbf{x}_{i+1}\Vert^2)+\eta(\Vert\mathbf{x}_i^\circ-\mathbf{x}_{i}\Vert^2-\Vert\mathbf{x}_i^\circ-\mathbf{x}_{i+1}\Vert^2)\label{EQ:THM-REd-1}
\end{align}
\normalsize
where $(a)$ follows from Lemma \ref{LM-StronglyConvex}, and $(b)$ is because
$\mathbf{Q}_{i+1}+\gamma{T}_i\mathbf{g}(\mathbf{x}_i)\succeq\mathbf{0}$ in
(\ref{EQ:VQ2}) and $\mathbf{g}(\mathbf{x}_i^\circ)\preceq\mathbf{0}$ in (\ref{EQ:xopt})
such that $[\mathbf{Q}_{i+1}+\gamma{T}_i\mathbf{g}(\mathbf{x}_i)]^T[\gamma{T}_{i+1}\mathbf{g}(\mathbf{x}_i^\circ)]\preceq\mathbf{0}$
for any $i\in\mathcal{I}$.

We now bound the RHS of (\ref{EQ:THM-REd-1}). Note that the aggregated loss
function $\frac{T_i}{S_i}\sum_{s\in\mathcal{S}_i}{f}_{\tau_i^s}(\mathbf{x})$
is $T_i\varrho$-strongly convex for any $i\in\mathcal{I}$.  Applying Lemma
\ref{LM-MGD} to the update of $\tilde{\mathbf{x}}_{i}^j$ in (\ref{eq:xij}),
for any $\alpha\ge{T}_iL$, we have
\small
\begin{align*}
        \Vert\mathbf{x}_i^\circ-\tilde{\mathbf{x}}_{i}^j\Vert^2\le\frac{\alpha-T_i\varrho}{\alpha+T_i\varrho}\Vert\mathbf{x}_i^\circ-\tilde{\mathbf{x}}_{i}^{j-1}\Vert^2,\quad\forall{j}\in\mathcal{J}.
\end{align*}
\normalsize
Note that the strong convexity constant $\varrho$ is smaller than the constant
of gradient Lipschitz continuity, \ie $\varrho\le{L}$ \cite{Mokhtari16}.
Combining the above $J$ inequalities and noting that $\tilde{\mathbf{x}}_i^0=\mathbf{x}_i$
by initialization, $1\le{T}_i\le{T}_{\text{max}}$, and $\alpha\ge{T}_{\text{max}}L$
such that $\frac{\alpha-T_i\varrho}{\alpha+T_i\varrho}\le\rho=\frac{\alpha-\varrho}{\alpha+\varrho}$,
we have
\small
\begin{align}
        \Vert\mathbf{x}_i^\circ-\tilde{\mathbf{x}}_{i}^{J}\Vert^2\le\rho^J\Vert\mathbf{x}_i^\circ-\mathbf{x}_{i}\Vert^2.\label{EQ:THM-REd-2}
\end{align}
\normalsize

Also,  from $\Vert\mathbf{a}+\mathbf{b}\Vert^2\ge\Vert\mathbf{a}\Vert^2+\Vert\mathbf{b}\Vert^2-2\Vert\mathbf{a}\Vert\Vert\mathbf{b}\Vert$
and the bound on $\mathcal{X}_0$ in (\ref{EQ:R}), we have
\small
\begin{align}
        &\Vert\mathbf{x}_i^\circ-\mathbf{x}_i\Vert^2-\Vert\mathbf{x}_i^\circ-\mathbf{x}_{i+1}\Vert^2\notag\\
        &\qquad\le\Vert\mathbf{x}_i^\circ-\mathbf{x}_i\Vert^2-\Vert\mathbf{x}_{i+1}^\circ-\mathbf{x}_{i+1}\Vert^2-\Vert\mathbf{x}_{i+1}^\circ-\mathbf{x}_i^\circ\Vert^2+2\Vert\mathbf{x}_{i+1}^\circ-\mathbf{x}_{i+1}\Vert\Vert\mathbf{x}_{i+1}^\circ-\mathbf{x}_i^\circ\Vert\le\Psi_i+2R\psi_i\label{EQ:THM-REd-3}
\end{align}
\normalsize
where we define $\Psi_i\triangleq\Vert\mathbf{x}_i^\circ-\mathbf{x}_i\Vert^2-\Vert\mathbf{x}_{i+1}^\circ-\mathbf{x}_{i+1}\Vert^2$
and $\psi_i\triangleq\Vert\mathbf{x}_i^\circ-\mathbf{x}_{i+1}^\circ\Vert$.

Substituting (\ref{EQ:THM-REd-2}) and (\ref{EQ:THM-REd-3}) into (\ref{EQ:THM-REd-1}),
we have
\small
\begin{align}
        &\frac{T_i}{S_i}\sum_{s\in\mathcal{S}_i}[\nabla{f}_{\tau_i^s}(\tilde{\mathbf{x}}_i^J)]^T(\mathbf{x}_{i+1}-\tilde{\mathbf{x}}_i^J)+\alpha\Vert\mathbf{x}_{i+1}-\tilde{\mathbf{x}}_i^{J}\Vert^2+[\mathbf{Q}_{i+1}+\gamma{T}_i\mathbf{g}(\mathbf{x}_i)]^T[\gamma{T}_{i+1}\mathbf{g}(\mathbf{x}_{i+1})]+\eta\Vert\mathbf{x}_{i+1}-\mathbf{x}_{i}\Vert^2\notag\\
        &\qquad\le\frac{T_i}{S_i}\!\sum_{s\in\mathcal{S}_i}\![\nabla{f}_{\tau_i^s}(\mathbf{x}_i)]^T(\mathbf{x}_i^\circ\!-\!\mathbf{x}_i)\!+\!(\alpha\rho^J\!+\!\eta)(\Psi_i\!+\!2R\psi_i).\!\!\label{EQ:THM-REd-4}
\end{align}
\normalsize

Adding $\frac{T_i}{S_i}\sum_{s\in\mathcal{S}_i}{f}_{\tau_i^s}(\mathbf{x}_i)$
on both sides of (\ref{EQ:THM-REd-4}), noting that ${f}_{\tau_i^s}(\mathbf{x}_i)+[\nabla{f}_{\tau_i^s}(\mathbf{x}_i)]^T(\mathbf{x}_i^\circ-\mathbf{x}_i)\le{f}_{\tau_i^s}(\mathbf{x}_i^\circ)$
from the convexity of $f_{\tau_i^s}(\mathbf{x})$ for any $s\in\mathcal{S}_i$,
and rearranging terms, we have
\small
\begin{align}
        &\frac{T_i}{S_i}\sum_{s\in\mathcal{S}_i}\left[f_{\tau_i^s}(\mathbf{x}_i)-f_{\tau_i^s}(\mathbf{x}_i^\circ)\right]\notag\\
        &\qquad\le-\frac{T_i}{S_i}\sum_{s\in\mathcal{S}_i}[\nabla{f}_{\tau_i^s}(\tilde{\mathbf{x}}_i^J)]^T(\mathbf{x}_{i+1}-\tilde{\mathbf{x}}_i^J)-\alpha\Vert\mathbf{x}_{i+1}-\tilde{\mathbf{x}}_i^{J}\Vert^2-[\mathbf{Q}_{i+1}+\gamma{T}_i\mathbf{g}(\mathbf{x}_i)]^T[\gamma{T}_{i+1}\mathbf{g}(\mathbf{x}_{i+1})]\notag\\
        &\qquad\quad-\eta\Vert\mathbf{x}_{i+1}-\mathbf{x}_{i}\Vert^2+(\alpha\rho^J+\eta)(\Psi_i+2R\psi_i).\label{EQ:THM-REd-5}
\end{align}
\normalsize

We then bound the RHS of (\ref{EQ:THM-REd-5}). Completing the
square, we have
\small
\begin{align}
        &-\frac{T_i}{S_i}\sum_{s\in\mathcal{S}_i}[\nabla{f}_{\tau_i^s}(\tilde{\mathbf{x}}_i^J)]^T(\mathbf{x}_{i+1}-\tilde{\mathbf{x}}_i^J)-\alpha\Vert\mathbf{x}_{i+1}-\tilde{\mathbf{x}}_i^{J}\Vert^2\notag\\
        &\qquad=-\frac{T_i}{S_i}\sum_{s\in\mathcal{S}_i}\left(\left\Vert\frac{\nabla{f}_{\tau_i^s}(\tilde{\mathbf{x}}_i^J)}{2\sqrt{\frac{\alpha}{T_i}}}+\sqrt{\frac{\alpha}{T_i}}(\mathbf{x}_{i+1}-\tilde{\mathbf{x}}_i^J)\right\Vert^2-\frac{T_{i}\left\Vert\nabla{f}_{\tau_i^s}(\mathbf{x}_i)\right\Vert^2}{4\alpha}\right)\stackrel{(a)}{\le}\frac{D^{2}T_{\text{max}}}{4\alpha}T_i.\label{EQ:THM-REd-6}
\end{align}
\normalsize
where $(a)$ follows by noting that $\nabla{f}_t(\mathbf{x})$ is bounded in
(\ref{EQ:D}).

Also, note that
\small
\begin{align}
        &-[\mathbf{Q}_{i+1}+\gamma{T}_i\mathbf{g}(\mathbf{x}_i)]^T[\gamma\mathbf{g}{T}_{i+1}(\mathbf{x}_{i+1})]&\notag\\
        &\qquad\stackrel{(a)}{\le}-\Delta_{i+1}+\Vert\gamma{T}_{i+1}\mathbf{g}(\mathbf{x}_{i+1})\Vert^2-[\gamma{T}_i\mathbf{g}(\mathbf{x}_i)]^T[\gamma{T}_{i+1}\mathbf{g}(\mathbf{x}_{i+1})]\notag\\
        &\qquad\stackrel{(b)}{=}-\Delta_{i+1}+\frac{\gamma^2}{2}\Vert{T}_{i+1}\mathbf{g}(\mathbf{x}_{i+1})\Vert^2-\frac{\gamma^2}{2}\Vert{T}_i\mathbf{g}(\mathbf{x}_i)\Vert^2+\frac{\gamma^2}{2}\Vert{T}_i\mathbf{g}(\mathbf{x}_i)-{T}_{i+1}\mathbf{g}(\mathbf{x}_{i+1})\Vert^2\notag\\
        &\qquad\stackrel{(c)}{\le}-\Delta_{i+1}+\frac{\gamma^2}{2}\Phi_i+\beta^2\gamma^2T_i^2\Vert\mathbf{x}_{i+1}-\mathbf{x}_{i}\Vert^2+\gamma^2G^2(T_i-T_{i+1})^2\label{EQ:THM-REd-7}
\end{align}
\normalsize
where $(a)$ follows from rearranging terms of (\ref{EQ:Drift}) in Lemma~\ref{LM-Drift}
such that $-\mathbf{Q}_{i+1}^T[\gamma{T}_{i+1}\mathbf{g}(\mathbf{x}_{i+1})]\le-\Delta_{i+1}+\Vert\gamma{T}_{i+1}\mathbf{g}(\mathbf{x}_{i+1})\Vert^2$,
$(b)$ is because $\mathbf{a}^T\mathbf{b}=\frac{1}{2}(\Vert\mathbf{a}\Vert^2+\Vert\mathbf{b}\Vert^2-\Vert\mathbf{a}-\mathbf{b}\Vert^2)$,
and $(c)$ follows from defining $\Phi_i\triangleq\Vert{T}_{i+1}\mathbf{g}(\mathbf{x}_{i+1})\Vert^2-\Vert{T}_i\mathbf{g}(\mathbf{x}_i)\Vert^2$,
$\frac{1}{2}\Vert\mathbf{a}+\mathbf{b}\Vert^2\le\Vert\mathbf{a}\Vert^2+\Vert\mathbf{b}\Vert^2$,
and $\mathbf{g}(\mathbf{x})$ being Lipschitz continuous in (\ref{EQ:Beta})
and bounded in (\ref{EQ:G}) such that
\small
\begin{align*}
        &\frac{1}{2}\Vert{T}_i\mathbf{g}(\mathbf{x}_i)-{T}_{i+1}\mathbf{g}(\mathbf{x}_{i+1})\Vert^2\notag\\
        &\qquad\le\Vert{T}_i\mathbf{g}(\mathbf{x}_i)-{T}_i\mathbf{g}(\mathbf{x}_{i+1})\Vert^2+\Vert{T}_i\mathbf{g}(\mathbf{x}_{i+1})-{T}_{i+1}\mathbf{g}(\mathbf{x}_{i+1})\Vert^2\le\beta^2T_i^2\Vert\mathbf{x}_{i}-\mathbf{x}_{i+1}\Vert^2+G^2(T_i-T_{i+1})^2.
\end{align*}
\normalsize

Substituting (\ref{EQ:THM-REd-6}) and (\ref{EQ:THM-REd-7}) into (\ref{EQ:THM-REd-5}),
and noting that  $\eta\ge\beta^2\gamma^2T_{\text{max}}^2$, we have
\small
\begin{align}
        \frac{T_i}{S_i}\sum_{s\in\mathcal{S}_i}\left(f_{\tau_i^s}(\mathbf{x}_i)-f_{\tau_i^s}(\mathbf{x}_i^\circ)\right)\le\frac{D^2T_{\text{max}}}{4\alpha}T_i+(\alpha\rho^J+\eta)(\Psi_i+2R\psi_i)-\Delta_{i+1}+\frac{\gamma^2}{2}\Phi_i+\gamma^2G^2(T_i-T_{i+1})^2.\label{EQ:THM-REd-8}
\end{align}
\normalsize

Summing (\ref{EQ:THM-REd-8}) over $i\in\mathcal{I}$, we have
\small
\begin{align}
        &\text{RE}_{\text{d}}(T)=\sum_{i\in\mathcal{I}}\frac{T_i}{S_i}\sum_{s\in\mathcal{S}_i}\left(f_{\tau_i^s}(\mathbf{x}_i)-f_{\tau_i^s}(\mathbf{x}_i^\circ)\right)\notag\\
        &\stackrel{(a)}{\le}\frac{D^2T_{\text{max}}}{4\alpha}T+(\alpha\rho^J+\eta)\left(\Vert\mathbf{x}_0^\circ-\mathbf{x}_0\Vert^2+2{R}\sum_{i\in\mathcal{I}}\Vert\mathbf{x}_i^\circ-\mathbf{x}_{i+1}^\circ\Vert\right)+L_1+\frac{\gamma^2}{2}\Vert{T}_I\mathbf{g}(\mathbf{x}_I)\Vert^2+\gamma^2G^2\sum_{i\in\mathcal{I}}(T_i-T_{i+1})^2\notag\\
        &\stackrel{(b)}{\le}\frac{D^2T_{\text{max}}}{4\alpha}T+(\alpha\rho^J+\eta)\left(R^2+2R\Pi_{\mathbf{x}^\circ}\right)+\gamma^2G^2(T_{\text{max}}^2+\Pi_T)\label{EQ:THM-REd-9}
\end{align}
\normalsize
where $(a)$ follows by noting that  $\Psi_i$, $\Delta_{i+1}$, and $\Phi_i$
are telescoping terms, such that their sums over $i\in\mathcal{I}$ are upper
bounded by $\Vert\mathbf{x}_0^\circ-\mathbf{x}_0\Vert^2$, $L_1$, and $\Vert{T}_I\mathbf{g}(\mathbf{x}_I)\Vert^2$,
respectively; $(b)$ follows from $\mathcal{X}_0$ being bounded in (\ref{EQ:R}),
$L_1=\frac{1}{2}\Vert\mathbf{Q}_1\Vert^2=\frac{1}{2}\Vert\gamma\mathbf{g}(\mathbf{x}_0)T_0\Vert^2\le\frac{1}{2}\gamma^2G^2T_{\text{max}}^2$,
$\Vert{T}_I\mathbf{g}(\mathbf{x}_I)\Vert^2\le{G}^2T_{\text{max}}^2$, and
the definitions of $\Pi_{\mathbf{x}^\circ}$ in (\ref{EQ:PL}) and $\Pi_{T}$
in (\ref{EQ:PiT}).\hfill$\blacksquare$
\vspace*{.6em}

The dynamic regret bound (\ref{EQ:RE-BD}) in Theorem \ref{THM-RE} improves as $J$ increases. When $J$ is large enough, we provide another dynamic regret bound for PQGA below.
\begin{theorem}\label{THM-REStrong}
For $J$ satisfying $2\rho^{J+1}<1$, if we choose $\alpha\ge{T_{\text{max}}}L$, $\eta\ge\max\{4\alpha,\beta^2\gamma^2T_{\text{max}}^2\}$, and $\gamma>0$, the dynamic regret of PQGA is upper bounded for any $\xi>0$ by
\begin{align}
        \text{RE}_{\text{d}}(T)&\le\frac{1}{4\xi}\Pi_{\nabla}\!+\!\frac{L+\xi}{1-2\rho^{J+1}}\left(R^2\!+\!\frac{1}{\alpha\!+\!\varrho}[\gamma^2G^2(T_{\text{max}}^2+\Pi_T)+\eta{R}(R+2\Pi_{\mathbf{x}^\circ})]\right)\label{EQ:RE-MGD}
\end{align}
where $\Pi_{\nabla}\triangleq\sum_{i\in\mathcal{I}}\Vert\frac{T_i}{S_i}\sum_{s\in\mathcal{S}_i}\nabla{f}_{\tau_i^s}(\mathbf{x}_i^\circ)\Vert^2$
is the accumulated squared gradients.
\end{theorem}
\textit{Proof:} We have
\small
\begin{align}
        \text{RE}_{\text{d}}(T)&=\sum_{i\in\mathcal{I}}\frac{T_i}{S_i}\sum_{s\in\mathcal{S}_i}\left(f_{\tau_i^s}(\mathbf{x}_i)-f_{\tau_i^s}(\mathbf{x}_i^\circ)\right)\stackrel{(a)}{\le}\sum_{i\in\mathcal{I}}\frac{T_i}{S_i}\sum_{s\in\mathcal{S}_i}[\nabla{f}_{\tau_i^s}(\mathbf{x}_i^\circ)]^T(\mathbf{x}_i-\mathbf{x}_i^\circ)+L\Vert\mathbf{x}_i-\mathbf{x}_i^\circ\Vert^2\notag\\
        &\stackrel{(b)}{\le}\frac{1}{4\xi}\Pi_{\nabla}+(L+\xi)\sum_{i\in\mathcal{I}}\Vert\mathbf{x}_i-\mathbf{x}_i^\circ\Vert^2\label{EQ:THM-MGD-1}
\end{align}
\normalsize
where $(a)$ follows from $f_t(\mathbf{x})$ being $2L$-smooth in (\ref{EQ:L}),
and $(b)$ is because $\mathbf{a}^T\mathbf{b}\le\frac{1}{4\xi}\Vert\mathbf{a}\Vert^2+\xi\Vert\mathbf{b}\Vert^2$
for any $\xi>0$ and  $\Pi_\nabla$ being defined under (\ref{EQ:RE-MGD}).

We now bound $\sum_{i\in\mathcal{I}}\Vert\mathbf{x}_i-\mathbf{x}_i^\circ\Vert^2$
on the RHS of (\ref{EQ:THM-MGD-1}). From  the inequality $\Vert\mathbf{a}+\mathbf{b}\Vert^2\le2(\Vert\mathbf{a}\Vert^2+\Vert\mathbf{b}\Vert^2)$
and  the bound on $\mathcal{X}_0$ in (\ref{EQ:R}), we have
\small
\begin{align}
        \sum_{i\in\mathcal{I}}\Vert\mathbf{x}_i-\mathbf{x}_i^\circ\Vert^2=\Vert\mathbf{x}_{0}-\mathbf{x}_{0}^\circ\Vert^2+\sum_{i\in\mathcal{I}}\Vert\mathbf{x}_{i+1}-\mathbf{x}_{i+1}^\circ\Vert^2\le{R}^2+2\sum_{i\in\mathcal{I}}\left(\Vert\mathbf{x}_{i+1}-\mathbf{x}_i^\circ\Vert^2+\Vert\mathbf{x}_i^\circ-\mathbf{x}_{i+1}^\circ\Vert^{2}\right).\label{EQ:THM-MGD-2}
\end{align}
\normalsize

We then bound $\sum_{i\in\mathcal{I}}\Vert\mathbf{x}_{i+1}-\mathbf{x}_i^\circ\Vert^2$
on the RHS of (\ref{EQ:THM-MGD-2}). From $f_{\tau_i^s}(\mathbf{x})$
being $2L$-smooth over $\mathcal{X}_0$
in (\ref{EQ:L}), we have
\small
\begin{align}
        f_{\tau_i^s}(\mathbf{x}_{i+1})&\le f_{\tau_i^s}(\tilde{\mathbf{x}}_{i}^J)+[\nabla{f}_{\tau_i^s}(\tilde{\mathbf{x}}_i^J)]^T(\mathbf{x}_{i+1}-\tilde{\mathbf{x}}_i^J)+L\Vert\mathbf{x}_{i+1}-\tilde{\mathbf{x}}_i^J\Vert^2,\quad\forall{s}\in\mathcal{S}_i.\label{EQ:THM-MGD-4}
\end{align}
\normalsize
From $f_{\tau_i^s}(\mathbf{x})$ being $2\varrho$-strongly convex over $\mathcal{X}_0$
in (\ref{EQ:varrho}), we have
\small
\begin{align}
        f_{\tau_i^s}(\mathbf{x}_{i}^\circ)&\ge f_{\tau_i^s}(\tilde{\mathbf{x}}_i^J)+[\nabla{f}_{\tau_i^s}(\tilde{\mathbf{x}}_i^J)]^T(\mathbf{x}_{i}^\circ-\tilde{\mathbf{x}}_i^J)+\varrho\Vert\mathbf{x}_{i}^\circ-\tilde{\mathbf{x}}_i^J\Vert^2,\quad\forall{s}\in\mathcal{S}_i.\label{EQ:THM-MGD-5}
\end{align}
\normalsize
We can show that (\ref{EQ:THM-REd-1}) in the proof of Theorem \ref{THM-RE}
still holds. Applying (\ref{EQ:THM-MGD-4}) and (\ref{EQ:THM-MGD-5}) to the
LHS and RHS of (\ref{EQ:THM-REd-1}), respectively, and rearranging terms,
we have
\small
\begin{align}
        \alpha\Vert\mathbf{x}_{i+1}-\mathbf{x}_i^\circ\Vert^2&\le\frac{T_i}{S_i}\!\sum_{s\in\mathcal{S}_i}\left(f_{\tau_i^s}(\mathbf{x}_{i}^\circ)-f_{\tau_i^s}(\mathbf{x}_{i+1})\right)-[\mathbf{Q}_{i+1}+\gamma{T}_i\mathbf{g}(\mathbf{x}_{i})]^T[\gamma{T}_{i+1}\mathbf{g}(\mathbf{x}_{i+1})]-\eta\Vert\mathbf{x}_{i+1}-\mathbf{x}_i\Vert^2\notag\\
        &\quad+\eta(\Vert\mathbf{x}_i-\mathbf{x}_i^\circ\Vert^2-\Vert\mathbf{x}_{i+1}-\mathbf{x}_i^\circ\Vert^2)+(\alpha-T_i\varrho)\Vert\tilde{\mathbf{x}}_i^J-\mathbf{x}_i^\circ\Vert^2-(\alpha-T_iL)\Vert\mathbf{x}_{i+1}-\tilde{\mathbf{x}}_i^J\Vert^2.\label{EQ:THM-MGD-6}
\end{align}
\normalsize

We now bound the right-hand side of (\ref{EQ:THM-MGD-6}).  Noting that $\frac{T_i}{S_i}\sum_{s\in\mathcal{S}_i}f_{\tau_i^s}(\mathbf{x})$
is $2T_i\varrho$-strongly convex over $\mathcal{X}_0$, from the definition
of $\mathbf{x}_i^\circ$ in (\ref{EQ:xtopt}), and applying Lemma \ref{LM-StronglyConvex}
again, we have
\small
\begin{align}
        \frac{T_i}{S_i}\sum_{s\in\mathcal{S}_i}f_{\tau_i^s}(\mathbf{x}_i^\circ)\le\frac{T_i}{S_i}\sum_{s\in\mathcal{S}_i}f_{\tau_i^s}(\mathbf{x}_{i+1})-T_i\varrho\Vert\mathbf{x}_{i+1}-\mathbf{\!x}_i^\circ\Vert^2.\label{EQ:THM-MGD-7}
\end{align}
\normalsize
We can show that (\ref{EQ:THM-REd-3}) and (\ref{EQ:THM-REd-7}) in the proof
of Theorem \ref{THM-RE} still hold. Substituting (\ref{EQ:THM-REd-3}), (\ref{EQ:THM-REd-7}),
and (\ref{EQ:THM-MGD-7}) into the RHS of (\ref{EQ:THM-MGD-6}) and rearranging
terms, we have
\small
\begin{align}
        &(\alpha+T_i\varrho)\Vert\mathbf{x}_{i+1}-\mathbf{x}_i^\circ\Vert^2\notag\\
        &\le-(\alpha-T_iL)\Vert\mathbf{x}_{i+1}-\tilde{\mathbf{x}}_i^J\Vert^2-(\eta-\beta^2\gamma^2T_i^2)\Vert\mathbf{x}_i-\mathbf{x}_{i+1}\Vert^2+(\alpha-T_i\varrho)\Vert\tilde{\mathbf{x}}_i^J\!-\!\mathbf{x}_i^\circ\Vert^2\notag\\
        &\quad-\eta\Vert\mathbf{x}_i^\circ-\mathbf{x}_{i+1}^\circ\Vert^2-\Delta_{i+1}+\frac{\gamma^2}{2}\Phi_i+\gamma^2G^2(T_i-T_{i+1})^2+\eta\Psi_i+2R\eta\Vert\mathbf{x}_i^\circ-\mathbf{x}_{i+1}^\circ\Vert\notag\\
        &\stackrel{(a)}{\le}(\alpha-T_i\varrho)\rho^J\Vert\mathbf{x}_i\!-\!\mathbf{x}_i^\circ\Vert^2-\eta\Vert\mathbf{x}_i^\circ\!-\!\mathbf{x}_{i+1}^\circ\Vert^2-\Delta_{i+1}+\frac{\gamma^2}{2}\Phi_i+\gamma^2G^2(T_i\!-\!T_{i+1})^2+\eta\Psi_i+2R\eta\Vert\mathbf{x}_i^\circ\!-\!\mathbf{x}_{i+1}^\circ\Vert\!\label{EQ:THM-MGD-10}
\end{align}
\normalsize
where $(a)$ follows from $\alpha\ge{T}_{\text{max}}L$ and $\eta\ge\beta^2\gamma^2T_{\text{max}}^2$,
and (\ref{EQ:THM-REd-2}) in the proof of Theorem \ref{THM-RE}.

Dividing both sides of (\ref{EQ:THM-MGD-10}) by $\alpha+T_i\varrho$ and substituting
it into (\ref{EQ:THM-MGD-2}), we have
\small
\begin{align}
        \sum_{i\in\mathcal{I}}\Vert\mathbf{x}_i-\mathbf{x}_i^\circ\Vert^2&\le{R}^2-\sum_{i\in\mathcal{I}}\left(\frac{\eta}{\alpha+T_i\varrho}-2\right)\Vert\mathbf{x}_i^\circ-\mathbf{x}_{i+1}^\circ\Vert^2+2\sum_{i\in\mathcal{I}}\rho^{J+1}\Vert\mathbf{x}_i-\mathbf{x}_i^\circ\Vert^2\notag\\
        &\quad+\sum_{i\in\mathcal{I}}\frac{1}{\alpha+T_i\varrho}\left(-\Delta_{i+1}+\frac{\gamma^2}{2}\Phi_i+\gamma^2G^2(T_i-T_{i+1})^2+\eta\Psi_i+2R\eta\Vert\mathbf{x}_i^\circ-\mathbf{x}_{i+1}^\circ\Vert\right)\notag\\
        &\stackrel{(a)}{\le}R^2+2\rho^{J+1}\sum_{i\in\mathcal{I}}\Vert\mathbf{x}_i-\mathbf{x}_i^\circ\Vert^2+\frac{1}{\alpha+\varrho}\left[\gamma^2G^2(T_{\text{max}}^2+\Pi_T)+\eta{R}(R+2\Pi_{\mathbf{x}^\circ})\right]\label{EQ:THM-MGD-11}
\end{align}
\normalsize
where $(a)$ follows from $\rho=\frac{\alpha-\varrho}{\alpha+\varrho}$, $\eta\ge4\alpha\ge2(\alpha+T_{\text{max}}\varrho)$,
and $(b)$ in the proof of (\ref{EQ:THM-REd-9}) for Theorem \ref{THM-RE}.

Substituting (\ref{EQ:THM-MGD-11}) into (\ref{EQ:THM-MGD-1}), on the condition
that $2\rho^{J+1}<1$, we have (\ref{EQ:RE-MGD}).\hfill$\blacksquare$

\subsubsection{Bounding the Static Regret}

Next, using the proof techniques for the dynamic regret $\text{RE}_{\text{d}}(T)$
in Theorem \ref{THM-RE}, we provide an upper bound on the static regret $\text{RE}_{\text{s}}(T)$
yielded by PQGA, given in the following theorem.

\begin{theorem}\label{THM-REs}
For any $J\ge0$, if we choose $\alpha\ge{T}_{\text{max}}L$, $\eta\ge\beta^2\gamma^2T_{\text{max}}^2$,
and $\gamma>0$, the static regret of PQGA is upper bounded by
\begin{align}
        \!\!\!\!\text{RE}_{\text{s}}(T)\!\le\!\frac{D^2T_{\text{max}}}{4\alpha}T\!+(\alpha\rho^J\!+\eta){R}^2+\gamma^2G^2(T_{\text{max}}^2\!+\Pi_T).\!\!\label{EQ:REs-BD}
\end{align}
\end{theorem}
\textit{Proof: (Proof outline)} Replacing all the per-period
optimizers $\{\mathbf{x}_i^\circ\}$ with the static offline benchmark $\mathbf{x}^\star$
in the proof of Theorem \ref{THM-RE}, we can show that for any $\alpha\ge{T}_{\text{max}}L$,
$\eta\ge\beta^2\gamma^2T_{\text{max}}$, and $\gamma>0$, (\ref{EQ:THM-REd-8})
still holds by redefining $\Psi_i\triangleq\Vert\mathbf{x}^\star-\mathbf{x}_i\Vert^2-\Vert\mathbf{x}^\star-\mathbf{x}_{i+1}\Vert^2$
and $\psi_i=0$. Summing the new version of (\ref{EQ:THM-REd-8}) over $i\in\mathcal{I}$,
and noting that $\Psi_i$ is still telescoping, we complete the proof.
\hfill$\blacksquare$

\subsubsection{Bounding the Constraint Violation}

We now proceed to provide an upper bound on the constraint violation $\text{VO}^c(T)$
for PQGA. We first relate the virtual queue vector $\mathbf{Q}_I$ to $\text{VO}^c(T)$
in the following lemma.

\begin{lemma}\label{LM-VOQ}
The periodic virtual queue vector yielded by PQGA satisfies the following
inequality: 
\begin{align}
        \text{VO}^c(T)\le\frac{1}{\gamma}\Vert\mathbf{Q}_I\Vert,\quad\forall{c}\in\mathcal{C}.\label{EQ:VOQ}
\end{align}
\end{lemma}
\textit{Proof:} From the periodic virtual queue dynamics in (\ref{EQ:VQ}),
for any $c\in\mathcal{C}$ and $i\in\mathcal{I}$, we have
\begin{align}
        {\gamma T}_i{g}^c(\mathbf{x}_i)\le{Q}_{i+1}^c-{Q}_i^c.\label{EQ:VOQ-1}
\end{align}
Summing (\ref{EQ:VOQ-1}) over $i\in\mathcal{I}$, we have
\small
\begin{align}
        \text{VO}^c(T)&=\sum_{i\in\mathcal{I}}{T}_ig^c(\mathbf{x}_i)\le\frac{1}{\gamma}\sum_{i\in\mathcal{I}}(Q_{i+1}^c-Q_i^c)=\frac{1}{\gamma}(Q_I^c-Q_0^c)\stackrel{(a)}{=}\frac{1}{\gamma}Q_I^c\stackrel{(b)}{\le}\frac{1}{\gamma}\Vert\mathbf{Q}_I\Vert\label{EQ:VOQ-2}
\end{align}
\normalsize
where $(a)$ follows from $Q_0^c=0$ by initialization, and $(b)$ is because
$\Vert\mathbf{a}\Vert_\infty\le\Vert\mathbf{a}\Vert$.
\hfill$\blacksquare$
\vspace*{0.7em}

Using Lemma \ref{LM-VOQ}, we can bound the constraint violation $\text{VO}^c(T)$
through an upper bound on the virtual queue vector $\mathbf{Q}_I$. The result
is stated in the following theorem.

\begin{theorem}\label{THM-VOc}
For any $J\ge0$, if we choose $\alpha,\eta,\gamma>0$, the constraint violation of PQGA is upper bounded for any constraint $c\in\mathcal{C}$ by
\begin{align}
        \text{VO}^c(T)\le2{G}T_{\text{max}}+\frac{(\alpha+\eta){R}^2+DRT_{\text{max}}^2+2\gamma^2{G}^2T_{\text{max}}}{\epsilon\gamma^2}.\label{EQ:VOc}
\end{align}
\end{theorem}
\textit{Proof:} 
Since $\mathbf{x}_{i+1}$ is chosen to solve $\textbf{P2}'$,
for any $i\in\mathcal{I}$, we have
\small
\begin{align}
        &\frac{T_i}{S_i}\sum_{s\in\mathcal{S}_i}[\nabla{f}_{\tau_i^s}(\tilde{\mathbf{x}}_i^J)]^T(\mathbf{x}_{i+1}-\tilde{\mathbf{x}}_i^J)+\alpha\Vert\mathbf{x}_{i+1}-\tilde{\mathbf{x}}_i^J\Vert^2+[\mathbf{Q}_{i+1}+\gamma{T}_i\mathbf{g}(\mathbf{x}_i)]^T[\gamma{T}_{i+1}\mathbf{g}(\mathbf{x}_{i+1})]+\eta\Vert\mathbf{x}_{i+1}-\mathbf{x}_i\Vert^2\notag\\
        &\qquad\le\frac{T_i}{S_i}\sum_{s\in\mathcal{S}_i}[\nabla{f}_{\tau_i^s}(\tilde{\mathbf{x}}_i^J)]^T(\mathbf{x}'-\tilde{\mathbf{x}}_i^J)+\alpha\Vert\mathbf{x}'-\tilde{\mathbf{x}}_i^J\Vert^2+[\mathbf{Q}_{i+1}+\gamma{T}_i\mathbf{g}(\mathbf{x}_i)]^T[\gamma{T}_{i+1}\mathbf{g}(\mathbf{x}')]+\eta\Vert\mathbf{x}'-\mathbf{x}_i\Vert^2.\label{EQ:THM-VOc-1}
\end{align}
\normalsize
Note that
\small
\begin{align}
        [\mathbf{Q}_{i+1}+\gamma{T}_i\mathbf{g}(\mathbf{x}_i)]^T[\gamma{T}_{i+1}\mathbf{g}(\mathbf{x}')]&\stackrel{(a)}{\le}-\epsilon\gamma{T}_{i+1}[\mathbf{Q}_{i+1}+\gamma{T}_i\mathbf{g}(\mathbf{x}_i)]^T\mathbf{1}\stackrel{(b)}{\le}-\epsilon\gamma{T}_{i+1}\Vert\mathbf{Q}_{i+1}+{\gamma{T}}_i\mathbf{g}(\mathbf{x}_i)\Vert\notag\\
        &\stackrel{(c)}{\le}-\epsilon\gamma{T}_{i+1}(\Vert\mathbf{Q}_{i+1}\Vert-\Vert\gamma{T}_i\mathbf{g}(\mathbf{x}_i)\Vert)\label{EQ:THM-VOc-2}
\end{align}
\normalsize
where $(a)$ follows from $\mathbf{x}'$ being an interior point of
$\mathbf{g}(\mathbf{x})$ in (\ref{EQ:Epsilon}) and $\mathbf{Q}_{i+1}+{\gamma{T}}_i\mathbf{g}(\mathbf{x}_i)\succeq\mathbf{0}$
in (\ref{EQ:VQ2}), $(b)$ is because $\Vert\mathbf{a}\Vert\le\Vert\mathbf{a}\Vert_1$,
and $(c)$ follows from $\left|({\Vert\mathbf{a}\Vert-\Vert\mathbf{b}\Vert})\right|\le\Vert\mathbf{a}-\mathbf{b}\Vert$
and $\Vert\mathbf{Q}_{i+1}\Vert\ge\Vert\gamma{T}_i\mathbf{g}(\mathbf{x}_i)\Vert$
in (\ref{EQ:VQ3}).

Substituting (\ref{EQ:THM-VOc-2}) into (\ref{EQ:THM-VOc-1}), and rearranging
terms, we have
\small
\begin{align}
        \mathbf{Q}_{i+1}^T[\gamma{T}_{i+1}\mathbf{g}(\mathbf{x}_{i+1})]&\le-\epsilon\gamma{T}_{i+1}(\Vert\mathbf{Q}_{i+1}\Vert-\Vert\gamma{T}_i\mathbf{g}(\mathbf{x}_i)\Vert)+\alpha\Vert\mathbf{x}'-\tilde{\mathbf{x}}_i^J\Vert^2+\eta\Vert\mathbf{x}'-\mathbf{x}_i\Vert^2\notag\\
        &\quad+\frac{T_i}{S_i}\sum_{s\in\mathcal{S}_i}[\nabla{f}_{\tau_i^s}(\tilde{\mathbf{x}}_i^J)]^T(\mathbf{x}'-\mathbf{x}_{i+1})-[\gamma{T}_i\mathbf{g}(\mathbf{x}_i)]^T[\gamma{T}_{i+1}\mathbf{g}(\mathbf{x}_{i+1})]\notag\\
        &\stackrel{(a)}{\le}-\epsilon\gamma T_{i+1}\Vert\mathbf{Q}_{i+1}\Vert+\epsilon\gamma^2{T}_{i+1}\Vert{T}_i\mathbf{g}(\mathbf{x}_i)\Vert+\alpha\Vert\mathbf{x}'-\tilde{\mathbf{x}}_i^J\Vert^2+\eta\Vert\mathbf{x}'-\mathbf{x}_i\Vert^2\notag\\
        &\quad+\frac{T_i}{S_i}\sum_{s\in\mathcal{S}_i}\Vert\nabla{f}_{\tau_i^s}(\tilde{\mathbf{x}}_i^J)\Vert\Vert\mathbf{x}'-\mathbf{x}_{i+1}\Vert+\gamma^2\Vert{T}_i\mathbf{g}(\mathbf{x}_i)\Vert\Vert{T}_{i+1}\mathbf{g}(\mathbf{x}_{i+1})\Vert\notag\\
        &\stackrel{(b)}{\le}-\epsilon\gamma{T}_{i+1}\Vert\mathbf{Q}_{i+1}\Vert+\epsilon\gamma^2{G}T_{i+1}T_i+(\alpha+\eta){R}^2+DRT_{i}+\gamma^2G^2T_{i+1}T_i\label{EQ:THM-VOc-3}
\end{align}
\normalsize
where $(a)$ is because $|\mathbf{a}^T\mathbf{b}|\le\Vert\mathbf{a}\Vert\Vert\mathbf{b}\Vert$;
and $(b)$ follows from the bound on $\mathbf{g}(\mathbf{x})$ in (\ref{EQ:G}),
the bound on $\mathcal{X}_0$ in (\ref{EQ:R}), and the bound on $\nabla{f}_t(\mathbf{x})$
in (\ref{EQ:D}). From (\ref{EQ:Drift}) in Lemma \ref{LM-Drift}, we have
\small
\begin{align}
        \Delta_{i+1}\le\mathbf{Q}_{i+1}^T[\gamma{T}_{i+1}\mathbf{g}(\mathbf{x}_{i+1})]+\Vert\gamma{T}_{i+1}\mathbf{g}(\mathbf{x}_{i+1})\Vert^2\le\mathbf{Q}_{i+1}^T[\gamma{T}_{i+1}\mathbf{g}(\mathbf{x}_{i+1})]+\gamma^2G^2T_{i+1}^2.\label{EQ:THM-VOc-4}
\end{align}
\normalsize
Substituting (\ref{EQ:THM-VOc-3}) into (\ref{EQ:THM-VOc-4}), we have
\small
\begin{align}
        \Delta_{i+1}&\le-\epsilon\gamma{T}_{i+1}\Vert\mathbf{Q}_{i+1}\Vert+\epsilon\gamma^2{}{G}T_{i+1}T_i+(\alpha+\eta){R}^2+DRT_{i}+\gamma^2G^2T_{i+1}T_i+\gamma^2G^2T_{i+1}^2.\label{EQ:THM-VOc-5}
\end{align}
\normalsize

Noting that $1\le{T}_i\le{T}_{\text{max}}$ for any $i\in\mathcal{I}$, from
(\ref{EQ:THM-VOc-5}), the sufficient condition for $\Delta_{i+1}<0$ is
\small
\begin{align*}
        \Vert\mathbf{Q}_{i+1}\Vert>{\gamma G}T_{\text{max}}+\frac{(\alpha+\eta){R}^2+DRT_{\text{max}}^2+2\gamma^2{G}^2T_{\text{max}}}{\epsilon\gamma}.
\end{align*}
\normalsize
If the above inequality holds, we have $\Vert\mathbf{Q}_{i+2}\Vert\le\Vert\mathbf{Q}_{i+1}\Vert$,
\ie the virtual queue length decreases; otherwise, from the virtual queue
bound in (\ref{EQ:VQ4}), there is a maximum increase from $\Vert\mathbf{Q}_{i+1}\Vert$
to $\Vert\mathbf{Q}_{i+2}\Vert$ since $\Vert\mathbf{Q}_{i+2}\Vert\le\Vert\mathbf{Q}_{i+1}\Vert+\Vert\gamma\mathbf{g}(\mathbf{x}_{i+1})T_{i+1}\Vert\le\Vert\mathbf{Q}_{i+1}\Vert+{\gamma
G}T_{\text{max}}$. Therefore, the virtual queue vector $\mathbf{Q}_I$ is
upper bounded by
\small
\begin{align*}
        \Vert\mathbf{Q}_{I}\Vert\le2\gamma{G}T_{\text{max}}+\frac{(\alpha+\eta){R}^2+DRT_{\text{max}}^2+2\gamma^2{G}^2T_{\text{max}}}{\epsilon\gamma}.
\end{align*} 
\normalsize

Substituting the above inequality into (\ref{EQ:VOQ}) in Lemma \ref{LM-VOQ},
we have (\ref{EQ:VOc}). 
\hfill$\blacksquare$

\subsection{Discussion on the Performance Bounds}
\label{Sec:Discuss}

In this section, we discuss the regret and constraint violation bounds of PQGA. To describe the level of time variation  of the
dynamic benchmark and update periods, we define parameters $\nu\ge0$ and
$\delta\ge0$ such that
\begin{align}
\Pi_{\mathbf{x}^\circ}&=\mathcal{O}(T^\nu),\label{EQ:nu}\\
\Pi_T&=\mathcal{O}(T^\delta)\label{EQ:delta}.
\end{align}
We show below that suitable values of parameters $\alpha$, $\eta$, and $\gamma$ for PQGA depend on whether $\nu$ and $\delta$ are known. Furthermore, the regret and constraint violation bounds also depend on the number of aggregated gradient descent steps $J$. We summarize the performance bounds of PQGA in Tables \ref{tab:anyJ} and \ref{tab:largeJ}.

\begin{table*}[t]
\renewcommand{\arraystretch}{1.2}
\caption{Dynamic Regret, Static Regret, and Constraint Violation Bounds of
PQGA for Any $J$ ($\Pi_{\mathbf{x}^\circ}=\mathcal{O}(T^\nu)$ and $\Pi_T=\mathcal{O}(T^\delta)$)}
\label{tab:anyJ}
\vspace{-2mm}
\centering
\small
\begin{tabular}{|c|c|c|c|c|c|c|}\hline 
$\Pi_T=\mathcal{O}(1)$ &Know $\nu$ and $\delta$&$\text{RE}_{\text{d}}(T)$&$\text{RE}_{\text{s}}(T)$&$\text{VO}^c(T)$\\\hline\hline
No& Yes &$\mathcal{O}(\max\{T^\frac{1+\nu}{2},T^{\delta+\kappa}\})$&$\mathcal{O}(\max\{T^\frac{1}{2},T^{\delta+\kappa}\})$&$\mathcal{O}(T^{\frac{1}{2}-\kappa})$\\\hline
No& No &$\mathcal{O}(\max\{T^{\frac{1}{2}+\nu},T^{\delta}\})$&$\mathcal{O}(\max\{T^\frac{1}{2},T^{\delta}\})$&$\mathcal{O}(T^{\frac{1}{2}})$\\\hline
Yes& Yes &$\mathcal{O}(T^\frac{1+\nu}{2})$&$\mathcal{O}(T^\frac{1}{2})$&$\mathcal{O}(1)$\\\hline
Yes& No &$\mathcal{O}(T^{\frac{1}{2}+\nu})$&$\mathcal{O}(T^\frac{1}{2})$&$\mathcal{O}(1)$\\\hline
\end{tabular}
\normalsize
\end{table*}

\begin{table*}[t]
\renewcommand{\arraystretch}{1.2}
\caption{Improved Dynamic Regret and Constraint Violation Bounds of PQGA
for a Large
$J$ ($\Pi_{\mathbf{x}^\circ}=\mathcal{O}(T^\nu)$ and $\Pi_T=\mathcal{O}(T^\delta)$)}
\label{tab:largeJ}
\vspace{-2mm}
\centering
\small
\begin{tabular}{|c|c|c|c|c|c|c|}\hline 
$\Pi_T=\mathcal{O}(1)$&Know $\nu$ and $\delta$&$\text{RE}_{\text{d}}(T)$&$\text{VO}^c(T)$\\\hline\hline
No& No &$\mathcal{O}(\max\{T^\nu,T^\delta\})$&$\mathcal{O}(1)$\\\hline
Yes& No &$\mathcal{O}(T^\nu)$&$\mathcal{O}(1)$\\\hline
\end{tabular}
\normalsize
\vspace{-2mm}
\end{table*}

\subsubsection{Regret and Constraint Violation Bounds for General~$J$}

From Theorems \ref{THM-RE}, \ref{THM-REs}, and \ref{THM-VOc}, we can derive
the following two corollaries regarding the regret and constraint violation bounds
for any $J\ge0$. The results can be easily obtained by substituting the chosen parameters $\alpha$, $\eta$, and $\gamma$ into the general performance bounds in (\ref{EQ:RE-BD}), (\ref{EQ:REs-BD}), and (\ref{EQ:VOc}), and we omit the algebraic details to avoid redundancy.

\begin{corollary}\label{Cor-1}
(Algorithm parameters with knowledge of $\nu$ and $\delta$) 
Let $\gamma^2=T^\kappa$, where $\kappa\in[0,\frac{1}{2}]$ is some trade-off parameter that can be freely chosen and $\eta=\beta^2\gamma^2T_{\text{max}}^2$. Then, for any $J\ge0$, $\text{RE}_{\text{d}}(T)=\mathcal{O}(\max\{T^\frac{1+\nu}{2},T^{\delta+\kappa}\})$
if we choose $\alpha={T}_{\text{max}}LT^{\frac{1-\nu}{2}}$, and $\text{RE}_{\text{s}}(T)=\mathcal{O}(\max\{T^\frac{1}{2},T^{\delta+\kappa}\})$
if we choose $\alpha={T}_{\text{max}}LT^{\frac{1}{2}}$. In both cases, $\text{VO}^c(T)=\mathcal{O}(T^{\frac{1}{2}-\kappa})$. Therefore, for any $0\le\nu<1$ and $0\le\delta<1$, and any $\kappa$ such that $\kappa+\delta<1$, both the dynamic and static regrets are sublinear and the constraint violation are sublinear.
\end{corollary}

\begin{corollary}\label{Cor-2}
(Algorithm parameters without knowledge of $\nu$ or $\delta$)
Let $\alpha={T}_{\text{max}}LT^{\frac{1}{2}}$, $\eta=\beta^2\gamma^2T_{\text{max}}^2$, and $\gamma^2=1$ in PQGA. Then, for any $J\ge0$, $\text{RE}_{\text{d}}(T)=\mathcal{O}(\max\{T^{\frac{1}{2}+\nu},T^{\delta}\})$, $\text{RE}_{\text{s}}(T)=\mathcal{O}(\max\{T^\frac{1}{2},T^{\delta}\})$, and $\text{VO}^c(T)=\mathcal{O}(T^{\frac{1}{2}})$.
\end{corollary}

From Corollaries \ref{Cor-1} and \ref{Cor-2}, a sufficient condition for PQGA to yield sublinear regrets under periodic updates is that the system variation measures $\Pi_{\mathbf{x}^\circ}$ and $\Pi_T$  grow sublinearly over time.  In many online applications, the system tends to stabilize over time, leading to sublinear system variation and thus sublinear regrets.

\begin{remark}
Sublinearly of the system variation measures is necessary to have sublinear
dynamic regret for OCO \cite{OT}. Otherwise, if the system varies too fast
over time, no online algorithm can track it due to the lack of in-time information.
This can be seen from the dynamic regret bounds derived in \cite{Zinkevich},
\cite{E.C.Hall}\nocite{CDC}\nocite{Mokhtari16}\nocite{L.Zhang17}-\cite{Dixit19},
\cite{T.Chen}, \cite{X.Cao}, \cite{INFOCOM21} even under the standard per-time-slot
update setting.
\end{remark}

\begin{remark}
With unknown time horizon $T$, the standard doubling trick \cite{OLBK}, \cite{CDC}, \cite{LTC-HY} can be applied to extend PQGA into one that has similar regret
bounds.
\end{remark}

\subsubsection{ Improved Dynamic Regret Bound for Large $J$}

From Theorems \ref{THM-REStrong} and \ref{THM-VOc}, we can derive the following
corollaries regarding the dynamic regret and constraint violation bounds
for PQGA, when the number of aggregated gradient descent steps $J$ is large
enough.

\begin{corollary}\label{Cor-3}
Suppose $\Pi_\nabla=\mathcal{O}(T^\nu)$.\footnote{The accumulated squared
gradients $\Pi_\nabla$ can be very small \cite{L.Zhang17}. In particular,
we have $\Pi_\nabla=0$ if $\mathbf{x}_i^\circ$ is an interior point of $\mathcal{X}_0$
(or there is no short-term constraint), \ie $\sum_{s\in\mathcal{S}_i}\nabla{f}_{\tau_i^s}(\mathbf{x}_i^\circ)=\mathbf{0}$
for any $i\in\mathcal{I}$.} Let $\alpha={T_{\text{max}}}L$, $\eta=\max\{4\alpha,\beta^2\gamma^2T_{\text{max}}^2\}$,
and $\gamma=1$. Then, for any $J\ge0$ such that $2\rho^{J+1}<1$, $\text{RE}_{\text{d}}(T)=\mathcal{O}(\max\{T^\nu,T^\delta\})$
and $\text{VO}^c(T)=\mathcal{O}(1)$. Therefore, for any $0\le\nu<1$ and $0\le\delta<1$,
both the dynamic regret and the constraint violation are sublinear.
\end{corollary}

\begin{remark}
When $J$ is large enough, the optimal algorithm parameters of PQGA do not
require the knowledge of
the system variation measures, \ie $\nu$ and $\delta$,
or the total time horizon $T$.
\end{remark}

\subsubsection{A Special Case of Bounded $\Pi_T$}

The following corollaries provide the regret and constraint violation
bounds when the accumulated variation of update periods is upper bounded
by a constant, \ie $\Pi_T=\mathcal{O}(1)$. In particular, this includes the
case where the update periods are fixed over time. The results are obtained
by setting $\delta=0$ in Corollaries \ref{Cor-1}-\ref{Cor-3}, respectively.

\begin{corollary}\label{Cor-4}
(Algorithm parameters with knowledge of $\nu$ and $\delta$) Let $\gamma^2=T^{\frac{1}{2}}$ and $\eta=\beta^2\gamma^2T_{\text{max}}^2$. Then, for any $J\ge0$, $\text{RE}_{\text{d}}(T)=\mathcal{O}(T^\frac{1+\nu}{2})$ if we chose $\alpha={T}_{\text{max}}LT^{\frac{1-\nu}{2}}$, and $\text{RE}_{\text{s}}(T)=\mathcal{O}(T^{\frac{1}{2}})$ if we choose $\alpha={T}_{\text{max}}LT^{\frac{1}{2}}$. In both cases, $\text{VO}^c(T)=\mathcal{O}(1)$.
\end{corollary}

\begin{corollary}\label{Cor-5}
(Algorithm parameters without knowledge of $\nu$ or $\delta$)
Let $\alpha={T}_{\text{max}}LT^{\frac{1}{2}}$, $\eta=\beta^2\gamma^2T_{\text{max}}^2$, and $\gamma^2=T^{\frac{1}{2}}$ in PQGA. Then, for any $J\ge0$, $\text{RE}_{\text{d}}(T)=\mathcal{O}(T^{\frac{1}{2}+\nu})$, $\text{RE}_{\text{s}}(T)=\mathcal{O}(T^\frac{1}{2})$, and $\text{VO}^c(T)=\mathcal{O}(1)$.
\end{corollary}

\begin{corollary}\label{Cor-6} 
Suppose $\Pi_\nabla=\mathcal{O}(T^\nu)$. Let $\alpha={T_{\text{max}}}L$,
$\eta=\max\{4\alpha,\beta^2\gamma^2T_{\text{max}}^2\}$, and $\gamma=1$. Then,
for any $J\ge0$ such that $2\rho^{J+1}<1$, $\text{RE}_{\text{d}}(T)=\mathcal{O}(T^\nu)$
and $\text{VO}^c(T)=\mathcal{O}(1)$.
\end{corollary}

\begin{remark}
PQGA can be applied to the special case of per-time-slot updates \cite{Trade}\nocite{LTC-Toff}\nocite{X.CaoTau}\nocite{T.Chen}\nocite{LTC-HY}\nocite{Yu-SC}\nocite{X.Cao}-\cite{INFOCOM21}.
Its regret and constraint violation bounds are given by Corollaries \ref{Cor-4} and \ref{Cor-5}, with $T_{\text{max}}=1$. In this case, PQGA recovers the known best $\mathcal{O}(T^{\frac{1}{2}})$ static regret and $\mathcal{O}(1)$ constraint violation in \cite{LTC-HY}. Furthermore, while \cite{LTC-HY} does not provide a dynamic regret bound, here we show that PQGA achieves $\mathcal{O}(T^\frac{1+\nu}{2})$
dynamic regret.\footnote{The analysis in \cite{LTC-HY} is for convex loss functions. We can show that PQGA still yields  $\mathcal{O}(T^\frac{1+\nu}{2})$
dynamic regret, $\mathcal{O}(T^{\frac{1}{2}})$ static regret, and $\mathcal{O}(1)$
constraint violation for convex loss functions with some care of
technical details and hence is omitted.}
\end{remark}

\begin{remark}
By increasing $J$, the dynamic regret of PQGA improves from $\mathcal{O}(T^\frac{1+\nu}{2})$ in Corollary \ref{Cor-4} to $\mathcal{O}(T^\nu)$ in Corollary \ref{Cor-6}, while maintaining the $\mathcal{O}(1)$ constraint violation.  To the best of our knowledge, even under the standard per-time-slot update setting, no known constrained OCO algorithm has simultaneously achieved $\mathcal{O}(T^\nu)$ dynamic regret and $\mathcal{O}(1)$ constraint violation.
\end{remark}


\section{Application to Massive MIMO System with Multiple Service Providers}
\label{Sec:WNV}

In many wireless systems, CSI is only available after a series of channel
estimation, quantization, and feedback processes. This challenge is especially
acute with massive MIMO systems, where the channel state space is large and
the channel state  can fluctuate quickly over time. Online learning provides
the tools to solve a variety of problems in dynamic MIMO systems.
 
As an example to study the performance of PQGA in practical systems, we apply
it to online network virtualization with massive MIMO antennas, where multiple
service providers (SPs) simultaneously share all the antennas and channel
resources provided by an infrastructure provider (InP). Most of the existing
works on MIMO virtualization have focused on offline problems \cite{rp}\nocite{EE17}\nocite{CRAN}\nocite{WNVNOMA}\nocite{AA}-\cite{V5G}.
Furthermore, these works enforce strict physical isolation among the SPs,
which suffer from performance loss in comparison with the spatial isolation approach
in \cite{Wang20} and \cite{Globecom19}. The virtualization solutions in \cite{Wang20}
and \cite{Globecom19} are online but they are based on Lyapunov optimization
and require the current CSI. Furthermore, neither of them considers periodic
precoder updates, which are essential to practical LTE and 5G NR networks.

\subsection{Online Precoding-Based Massive MIMO Network Virtualization}

We consider an InP performing network virtualization in  a massive MIMO cellular
network. In each cell, the InP owns a base station (BS) equipped with $N$
antennas, serving $M$ SPs. Let $\mathcal{M}=\{1,\dots,M\}$.
 Each SP $m$ has $K_m$ users. Let the total number
of users in the cell be $K$. We consider a time-slotted system with time
indexed by $t$. Let $\mathbf{H}_t^m\in\mathbb{C}^{K_m\times{N}}$ be the local
CSI between the BS and the $K_m$ users of SP $m$ at time $t$.
\subsubsection{Precoding-Based Network Virtualization}

\begin{figure}[!t]
\centering
{\includegraphics[width=.5\linewidth,trim= 300 125 180 120,clip]{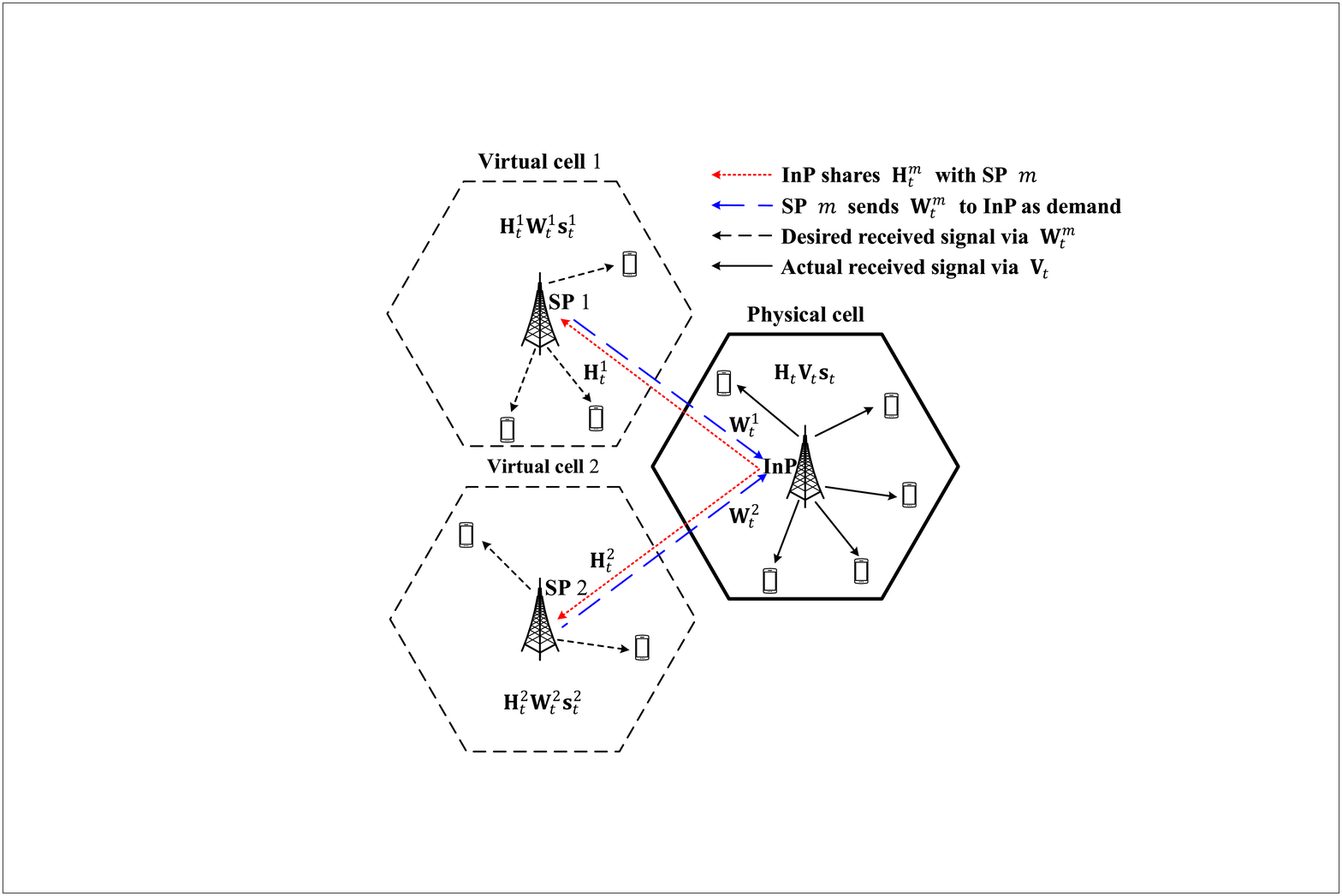}}
\vspace{-2mm}
\caption {An illustration of idealized massive MIMO virtualization in a cell
with one InP and two SPs serving users in their respective virtual cell.}
\label{Fig:CMIMO}
\vspace{-4mm}
\end{figure}

For ease of exposition, we first consider an idealized massive MIMO virtualization
framework, where CSI is fedback per time slot without delay, as shown in
Fig.~\ref{Fig:CMIMO}. At each time slot $t$, the InP shares the corresponding
local CSI $\mathbf{H}_t^{m}$ with SP $m$, and it allocates transmit power
$P_m$ to the SP. The power allocation is limited by the total transmit power
budget $P_{\text{max}}$, \ie $\sum_{m\in\mathcal{M}}P_m \le P_{\text{max}}$.
Using $\mathbf{H}_t^m$, each SP $m$ designs its own precoding matrix $\mathbf{W}_t^{m}\in\mathbb{C}^{N\times{K}_m}$
based on the service needs of its  users, while ensuring $\Vert\mathbf{W}_t^{m}\Vert_F^2\le{P}_m$.
The SP then sends $\mathbf{W}_t^m$ to the InP as its service demand.
Note that each SP $m$ designs $\mathbf{W}_t^{m}$ based only on its local
CSI, not needing to be aware of the users of the other SPs. For SP $m$, its
\textit{desired} received signal vector $\tilde{\mathbf{y}}_t^{m}$ at its
$K_m$ users is given by
\begin{align*}
        \tilde{\mathbf{y}}_t^{m}=\mathbf{H}_t^m\mathbf{W}_t^m\mathbf{s}_t^m,\quad\forall{m}\in\mathcal{M}
\end{align*}
where $\mathbf{s}_t^m$ is the transmitted signal vector from SP $m$ to its
$K_m$ users. Let $\tilde{\mathbf{y}}_t\triangleq[{\hbox{$\tilde{\mathbf{y}}_t^1$}}^H,\dots,{\hbox{$\tilde{\mathbf{y}}_t^M$}}^H]^H$
be the desired received signal vector at all $K$ users, $\mathbf{D}_t\triangleq
\blkdiag\{\mathbf{H}_t^1\mathbf{W}_t^1,\dots,\mathbf{H}_t^M\mathbf{W}_t^M\}$
be the virtualization demand made by the SPs, and $\mathbf{s}_t\triangleq[{\mathbf{s}_t^1}^H,\dots,{\mathbf{s}_t^M}^H]^H$.
Then we have $\tilde{\mathbf{y}}_t=\mathbf{D}_t\mathbf{s}_t$. We assume that
the transmitted signals to all $K$ users are independent to each other, with
$\mathbb{E}\{\mathbf{s}_t\mathbf{s}_t^H\}=\mathbf{I},\forall t$. 

At each time slot $t$, the InP has the global CSI $\mathbf{H}_t=[{\mathbf{H}_t^1}^H,\dots,{\mathbf{H}_t^M}^H]^H\in\mathbb{C}^{K\times{N}}$
and designs the \textit{actual} global downlink precoding matrix $\mathbf{V}_t\triangleq[\mathbf{V}_t^{1},\dots,\mathbf{V}_t^{M}]\in\mathbb{C}^{N\times{K}}$
to serve all $K$ users, where $\mathbf{V}_t^m\in\mathbb{C}^{N\times{K}_m}$
is the actual downlink precoding matrix for SP $m$. Then, the actual received
signal vector $\mathbf{y}_t^m$ at the users of SP $m$ is given by
\begin{align*}
        \mathbf{y}_t^m=\mathbf{H}_t^m\mathbf{V}_t^m\mathbf{s}_t^m+\sum_{l\neq{m},l\in\mathcal{M}}\mathbf{H}_t^m\mathbf{V}_t^l\mathbf{s}_t^l,\quad\forall{m}\in\mathcal{M}
\end{align*}
where the second term is the inter-SP interference from the other SPs to
the users of SP $m$. The actual received signal vector $\mathbf{y}_t\triangleq[{\mathbf{y}_t^1}^H,\dots,{\mathbf{y}_t^M}^H]^H$
at all $K$ users is given by $\mathbf{y}_t=\mathbf{H}_t\mathbf{V}_t\mathbf{s}_t$.

For downlink massive MIMO network virtualization, the InP designs the actual
precoding matrix $\mathbf{V}_t$ to mitigate the inter-SP interference in
order to meet the virtualization demand $\mathbf{D}_t$ received from the
SPs. The expected deviation of the actual received signals from that of the
SPs' virtualization demand is given by $\mathbb{E}\{\Vert\mathbf{y}_t-\tilde{\mathbf{y}}_t\Vert_F^2\}=\Vert\mathbf{H}_t\mathbf{V}_t-\mathbf{D}_t\Vert_F^2$.
Therefore, we define the \textit{precoding deviation} for any precoding matrix
$\mathbf{V}$ as follows:
\begin{align}
        f_{t}(\mathbf{V})\triangleq\Vert\mathbf{H}_{t}\mathbf{V}-\mathbf{D}_{t}\Vert_F^2,\quad\forall{t}\in\mathcal{T}\label{EQ:fV}
\end{align}
which we use as the design metric for massive MIMO network virtualization.
Note that $f_{t}(\mathbf{V})$ naturally quantifies the difference between
the actual global precoder executed at the BS and the virtual local precoders
demanded by the SPs. Furthermore, it is strongly convex in $\mathbf{V}$.

\subsubsection{Online Precoding Optimization with Periodic Updates}

Under the transmission structure of a typical cellular network, such as LTE
and 5G NR, we consider an online periodic virtualization demand-response
mechanism. An update period may correspond to the duration of one or multiple
resource blocks and can vary over time. Within each update period $i\in\mathcal{I}$,
the InP, which is the decision maker as defined in Section \ref{Sec:Problem},
receives multiple delayed CSI $\mathbf{H}_t$ and virtualization demand $\mathbf{D}_t$
feedbacks for $t\in\mathcal{T}_i$. At the beginning of each update period
$i$, the InP determines $\mathbf{V}_{i}$ in the  compact convex set 
\begin{align}
        \mathcal{V}_{0}\triangleq\{\mathbf{V}:\Vert\mathbf{V}\Vert_F^2\le{P}_{\text{max}}\}\label{EQ:STP}
\end{align}

\noindent to meet the short-term transmit power constraint. We also consider
a long-term transmit power constraint, with 
\begin{align}
        g(\mathbf{V})\triangleq\Vert\mathbf{V}\Vert_F^2-\bar{P}\label{EQ:gV}
\end{align}
being the long-term transmit power constraint function, where $\bar{P}\le{P}_{\text{max}}$
is the average transmit power budget.  

Then, our online optimization problem for MIMO network virtualization is
in the same form as $\textbf{P1}$, with the loss function, short-term constraint,
and long-term constraint function given in (\ref{EQ:fV}), (\ref{EQ:STP}),
and (\ref{EQ:gV}), respectively.

\subsection{Online Precoding Solution}
\label{Sec:Online Precoding Solution}

Using the proposed PQGA algorithm, at the beginning of each update period
$i+1$, we first initialize an intermediate precoder $\tilde{\mathbf{V}}_i^0=\mathbf{V}_{i}$.
If $J>0$, for each $j\in\mathcal{J}$, we solve the following precoder optimization
problem for $\tilde{\mathbf{V}}_i^{j}$:
\begin{align*}
        \min_{\mathbf{V}\in\mathcal{V}_{0}}~\frac{T_i}{S_i}\sum_{s\in\mathcal{S}_i}2\Re\{\tr\{[\nabla_{\tilde{\mathbf{V}}_i^{j-1*}}f_{\tau_i^s}(\tilde{\mathbf{V}}_i^{j-1})]^H(\mathbf{V}\!-\!\tilde{\mathbf{V}}_i^{j-1})\}\}+\alpha\Vert\mathbf{V}-\tilde{\mathbf{V}}_i^{j-1}\Vert_F^2
\end{align*}
where $\nabla_{\tilde{\mathbf{V}}_i^{j-1*}}f_{\tau_i^s}(\tilde{\mathbf{V}}_i^{j-1})=\mathbf{H}_{\tau_i^s}^H(\mathbf{H}_{\tau_i^s}\tilde{\mathbf{V}}_i^{j-1}-\mathbf{D}_{\tau_i^s})$.
This is equivalent to performing projected aggregated
gradient descent with a closed-form solution for $\tilde{\mathbf{V}}_i^j$,
given by
\begin{align}
\tilde{\mathbf{V}}_{i}^j=\left\{
\begin{matrix} 
        \tilde{\mathbf{X}}_i^j,&\text{if~}\Vert\tilde{\mathbf{X}}_i^j\Vert_F^2\leq{P}_{\text{max}}\\
        \sqrt{P_{\text{max}}}\frac{\tilde{\mathbf{X}}_i^j}{\Vert\tilde{\mathbf{X}}_i^j\Vert_F},&\text{o.w.}
\end{matrix}
\right.\label{EQ:closedJ}
\end{align}
where $\tilde{\mathbf{X}}_i^j=\tilde{\mathbf{V}}_i^{j-1}-\frac{T_i}{\alpha{S}_i}\sum_{s\in\mathcal{S}_i}\mathbf{H}_{\tau_i^s}^H(\mathbf{H}_{\tau_i^s}\tilde{\mathbf{V}}_i^{j-1}-\mathbf{D}_{\tau_i^s})$.

After performing $J$-step aggregated gradient descent, with both $\tilde{\mathbf{V}}^J$
and $\mathbf{V}_i$, we solve the following precoder optimization problem
for the actual precoding matrix $\mathbf{V}_{i+1}$:
\begin{align*}
        \textbf{P3}':~\min_{\mathbf{V}\in\mathcal{V}_{0}}~&\frac{T_i}{S_i}\sum_{s\in\mathcal{S}_i}2\Re\{\tr\{[\nabla_{\tilde{\mathbf{V}}_i^{J*}}f_{\tau_i^s}(\tilde{\mathbf{V}}_i^J)]^H(\mathbf{V}-\tilde{\mathbf{V}}_i^J)\}\}+\alpha\Vert\mathbf{V}-\tilde{\mathbf{V}}_i^J\Vert_F^2+\eta\Vert\mathbf{V}-\mathbf{V}_i\Vert_F^2\notag\\
        &\quad+[Q_{i+1}+\gamma{T}_{i}{g}(\mathbf{V}_{i})][\gamma{T}_{i+1}{g}(\mathbf{V})]
\end{align*}
where $Q_{i+1}$ is a periodic virtual queue with updating rule in (\ref{EQ:VQ}).
Since $\textbf{P3}'$ is a strongly convex optimization problem with strong
duality, we solve it by inspecting Karush-Kuhn-Tucker (KKT) conditions \cite{Boyd}.
The Lagrangian for $\textbf{P3}'$ is
\begin{align*}
        L(\mathbf{V},\lambda)&=\frac{T_i}{S_i}\sum_{s\in\mathcal{S}_i}2\Re\{\tr\{[\nabla_{\tilde{\mathbf{V}}_i^{J*}}f_{\tau_i^s}(\tilde{\mathbf{V}}_i^J)]^H(\mathbf{V}\!-\!\tilde{\mathbf{V}}_i^J)\}\}+\alpha\Vert\mathbf{V}\!-\!\tilde{\mathbf{V}}_i^J\Vert_F^2+\eta\Vert\mathbf{V}\!-\!\mathbf{V}_i\Vert_F^2\notag\\
        &\quad+[Q_{i+1}+\gamma{T}_i{g}(\mathbf{V}_i)][\gamma{T}_{i+1}{g}(\mathbf{V})]+\lambda(\Vert\mathbf{V}\Vert_F^{2}-P_{\text{max}})
\end{align*}
where $\lambda$ is the Lagrange multiplier associated with the short-term
transmit power constraint in (\ref{EQ:STP}). The KKT conditions for $(\mathbf{V}^{\star},\lambda^{\star})$
being globally optimal are given by $\Vert\mathbf{V}^{\star}\Vert_F^2-P_{\text{max}}\le0$,
$\lambda^{\star}\geq 0$, $\lambda^{\star}(\Vert\mathbf{V}^{\star}\Vert_F^2-P_{\text{max}})=0$,
and
\begin{align}
        \mathbf{V}^{\star}=\frac{\alpha\tilde{\mathbf{V}}_i^J+\eta\mathbf{V}_i-\frac{T_i}{S_i}\sum_{s\in\mathcal{S}_i}\mathbf{H}_{\tau_i^s}^H(\mathbf{H}_{\tau_i^s}\tilde{\mathbf{V}}_i^J-\mathbf{D}_{\tau_i^s})}{\alpha+\eta+[Q_{i+1}+{\gamma{T}}_i{g}(\mathbf{V}_i)]\gamma{T}_{i+1}+\lambda^{\star}},\label{EQ:P2KKT1}
\end{align}
which follows from setting $\nabla_{\mathbf{V}^*}L(\mathbf{V},\lambda)=\mathbf{0}$.
From  the KKT conditions, and noting that $\lambda^\star$ serves as a power
regularization factor for $\mathbf{V}^\star$ in (\ref{EQ:P2KKT1}), we have
a closed-form solution for $\mathbf{V}_{i+1}$, given by
\begin{align}
\mathbf{V}_{i+1}=\left\{
\begin{matrix} 
        \mathbf{X}_i,&\text{if~}\Vert\mathbf{X}_i\Vert_F^2\leq{P}_{\text{max}}\\
        \sqrt{P_{\text{max}}}\frac{\mathbf{X}_i}{\Vert\mathbf{X}_i\Vert_F},&\text{o.w.}
\end{matrix}
\right.\label{EQ:closed}
\end{align}
where $\mathbf{X}_i=\frac{\alpha\tilde{\mathbf{V}}_i^J+\eta\mathbf{V}_i-\frac{T_i}{S_i}\sum_{s\in\mathcal{S}_i}\mathbf{H}_{\tau_i^s}^H(\mathbf{H}_{\tau_i^s}\tilde{\mathbf{V}}_i^J-\mathbf{D}_{\tau_i^s})}{\alpha+\eta+[Q_{i+1}+\gamma{T}_i{g}(\mathbf{V}_i)]\gamma{T}_{i+1}}$.

The computational complexities of our online precoding solutions in (\ref{EQ:closedJ})
and (\ref{EQ:closed}) are dominated by matrix multiplications. They are in
the order of $\mathcal{O}(NK^2)$.

\begin{remark}
Note that additional short-term per-antenna transmit power constraints can
be incorporated in the convex set $\mathcal{V}_0$. In this case, both precoder optimization problems can be equivalently
decomposed into $N$ subproblems, each with a closed-form solution similar
to (\ref{EQ:closedJ}) and (\ref{EQ:closed}).
\end{remark}

\subsection{Performance Bounds}

We assume that the channel gain is bounded by a constant $B>0$ at any time
$t$, given by
\begin{align}
        \Vert\mathbf{H}_t\Vert_F\le{B},\quad\forall{t}\in\mathcal{T}.\label{EQ:B}
\end{align}
We show in the following lemma that our online MIMO network virtualization
problem satisfies the OCO Assumptions \ref{ASP-strong}-\ref{ASP-3} made in
Section \ref{Sec:Bounds}. The proof follows from the above bound on $\mathbf{H}_t$
and the short-term transmit power limit on both $\mathbf{V}_t$ and $\mathbf{W}_t$,
and is omitted for brevity.

\begin{lemma}\label{LM-BD}
Assume the bounded channel gain in (\ref{EQ:B}). Then, Assumptions \ref{ASP-strong}-\ref{ASP-3}
hold with the corresponding constants given by $\varrho=2$, $L=B^2$, $D=4B^2\sqrt{P_{\text{max}}}$,
$\beta=2\sqrt{P_{\text{max}}}$, $G=\sqrt{\max\{{\bar{P}}^2,(P_{\text{max}}-\bar{P})^2\}}$,
$\epsilon=\bar{P}$, and $R=2\sqrt{P_{\text{max}}}$.\end{lemma}

From the results in Theorems \ref{THM-RE}-\ref{THM-VOc}, the performance bounds yielded by $\{\mathbf{V}_i\}$ are given by (\ref{EQ:RE-BD})-(\ref{EQ:REs-BD}) and (\ref{EQ:VOc}), with the corresponding $\varrho,L,D,\beta,G,\epsilon,R$ given in Lemma~\ref{LM-BD}. 


\section{Simulation Results}
\label{Simulation}

In this section, we present simulation results for applying PQGA to online
precoding-based massive MIMO network virtualization, under typical urban
micro-cell LTE network settings. We study the impact of various system parameters
on algorithm convergence and performance. We numerically demonstrate the performance advantage of PQGA over the known best alternative.

\subsection{Simulation Setup}

We consider an urban hexagon micro-cell of radius $500$~m. An InP owns the
BS, which is equipped with $N=32$ antennas as default. The InP performs network
virtualization and serves $M=4$ SPs. We focus on some arbitrary channel over
one subcarrier with bandwidth $B_W=15$ kHz. Within this channel, each SP
$m\in\mathcal{M}$ serves $K_m=2$ users uniformly distributed in the cell,
for a total of $K=8$ users in the cell. We set maximum transmit power limit
$P_{\text{max}}=33$~dBm, time-averaged transmit power limit $\bar{P}=30$~dBm,
noise power spectral density $N_0=-174$ dBm/Hz, and noise figure $N_F=10$
dB, as default system parameters. 

We model the fading channel as a first-order Gaussian-Markov process  $\mathbf{h}_{t+1}^{k}=\alpha_{\mathbf{h}}\mathbf{h}_t^{k}+\mathbf{z}_t^{k},\forall{k}\in\mathcal{K}=\{1,\dots,K\}$,
where $\mathbf{h}_t^{k}\sim \mathcal{CN}(\mathbf{0},\beta_{k}\mathbf{I})$,
with $\beta_k{[\text{dB}]}= -31.54-33\log_{10}(d_k)-\psi_k$ capturing the
path-loss and shadowing, $d_k$ being the distance  from  the BS to user $k$,
$\psi_k\sim\mathcal{CN}(0,\sigma_{\phi}^2)$ accounting\ for the shadowing
effect with $\sigma_{\phi}= 8$ dB, $\alpha_\mathbf{h}\in[0,1]$ is the channel
correlation coefficient, and $\mathbf{z}_t^{k}\sim\mathcal{CN}(\mathbf{0},(1-\alpha_{\mathbf{h}}^2)\beta_{k}\mathbf{I})$
is independent of $\mathbf{h}_t^{k}$. We set $\alpha_\mathbf{h}=0.997$ as
default, which corresponds to pedestrian user speed $1$~m/s, under the standard
LTE transmission structure \cite{LTEP}. \footnote{We emphasize here that the Gauss-Markov channel model is used for illustration only. PQGA can be applied to any arbitrary wireless environment, and the InP does not need to know the channel statistics.} We set the time slot duration $\Delta{t}=\frac{1}{B_W}$, such that an update period of $8$ time slots is similar to one resource block time duration in LTE. We set the time horizon $T=400$.

We assume that each SP $m\in\mathcal{M}$ uses zero forcing (ZF)
precoding scheme $\varpi_t^m{\mathbf{H}_t^m}^H(\mathbf{H}_t^m{\mathbf{H}_t^m}^H)^{-1}$
to design its virtual precoding matrix $\mathbf{W}_t^m$, where $\varpi_t^m$
is a power normalizing factor such that $\Vert\mathbf{W}_t^m\Vert_F^2=P_m=\frac{P_{\text{max}}}{M}$. For performance evaluation, we define the time-averaged precoding deviation normalized against the virtualization demand as
\begin{align*}
        \bar{f}(T)\triangleq\frac{1}{T}\sum_{i\in\mathcal{I}}\sum_{t\in\mathcal{T}_i}\frac{f_t(\mathbf{V}_i)}{\Vert\mathbf{D}_t\Vert_F^2},
\end{align*}
the time-averaged transmit power as
\begin{align*}
        \bar{P}(T)\triangleq\frac{1}{T}\sum_{i\in\mathcal{I}}T_i\Vert\mathbf{V}_i\Vert_F^2,
\end{align*}
and the time-averaged per-user rate as
\begin{align*}
        \bar{R}(T)\triangleq\frac{1}{TK}\sum_{i\in\mathcal{I}}\sum_{t\in\mathcal{T}_i}\sum_{k\in\mathcal{K}}\log_2\left(1+\text{SINR}_t^{ik}\right)
\end{align*}
where $\text{SINR}_t^{ik}=\frac{|{\mathbf{h}_t^k}^T\mathbf{v}_i^k|^2}{\sum_{j\neq{k},j\in\mathcal{K}}|{\mathbf{h}_t^k}^T\mathbf{v}_i^j|^2+\sigma_n^2}$,
with $\mathbf{h}_t^k$ and $\mathbf{v}_i^k$ being the channel vector at time
slot $t$ and precoding vector in the $i$-th update period for user $k$, and
$\sigma_n^2=N_0B_W+N_F$ being the noise power. 

\subsection{Impact of Update Periods}

We first consider the update periods $\{T_i\}$ are fixed over time and there
is only one CSI feedback at the beginning of each update period $i$. Fig.~\ref{fig:Ti}
shows $\bar{f}(T)$, $\bar{P}(T)$, and $\bar{R}(T)$ versus $T$ with different
values of the update period $T_i$. We observe that PQGA converges fast. The performance of PQGA deteriorates as $T_i$ increases, \ie precoder updates are less frequent. This illustrates the impact of channel variation over time, while the precoder is fixed. The long-term transmit power $\bar{P}(T)$ quickly converges to the average transmit power limit $\bar{P}$.

\subsection{Impact of Number of Aggregated Gradient Descent Steps}
\label{Sec:Impact of Number of Aggregated Gradient Descent Steps}

\begin{figure}[!t]
\centering
\vspace{-2mm}
\begin{minipage}{0.495\textwidth}
\hspace{7.3mm}
\includegraphics[width=.845\linewidth,trim= 0 35 00 30,clip]{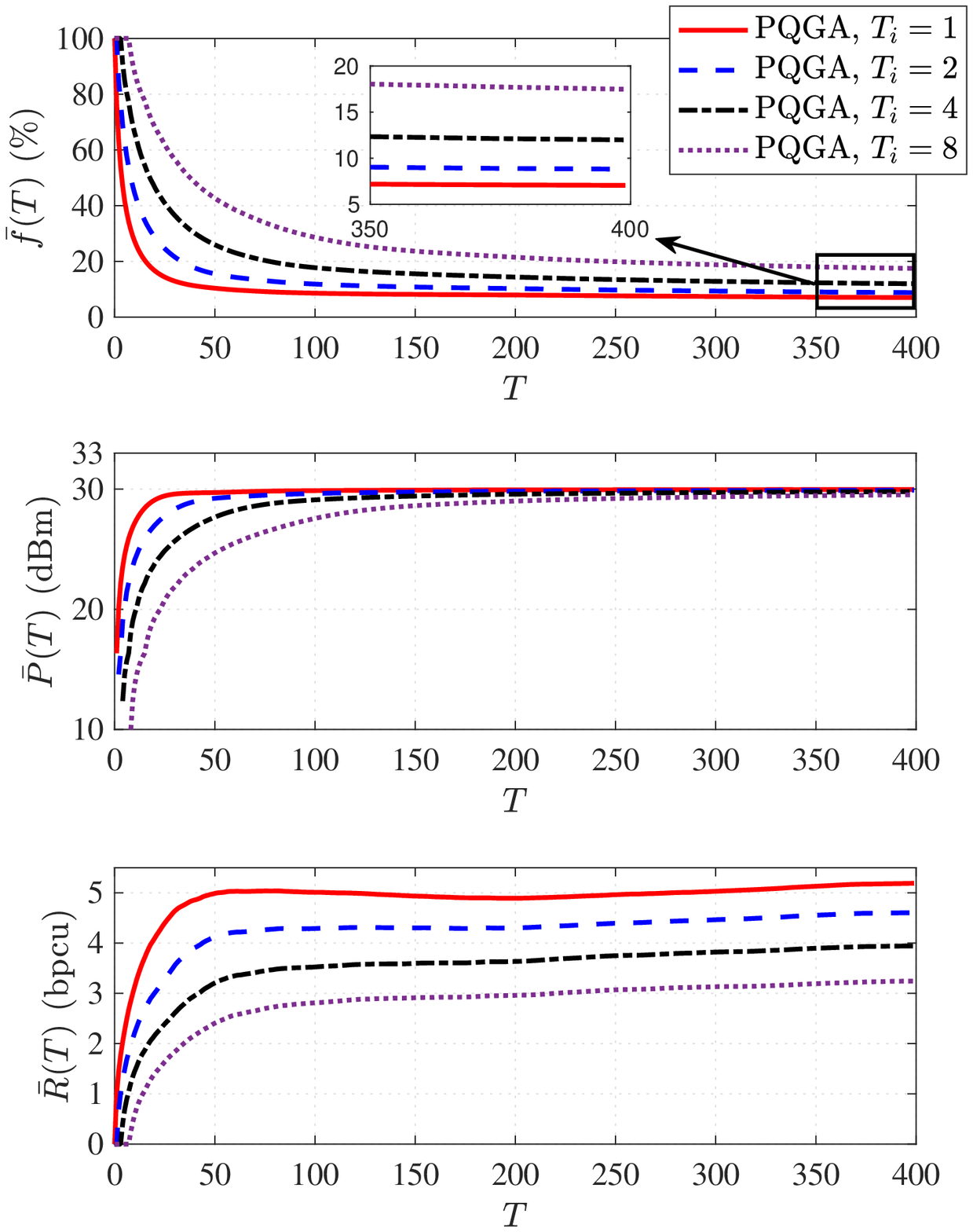}
\vspace{-4mm}
\caption{$\bar{f}(T)$, $\bar{P}(T)$, and $\bar{R}(T)$ vs. $T$ with different
$T_i$ values.}\label{fig:Ti}
\end{minipage}
\hfill
\begin{minipage}{0.495\textwidth}
\includegraphics[width=.845\linewidth,trim= 0 35 00 30,clip]{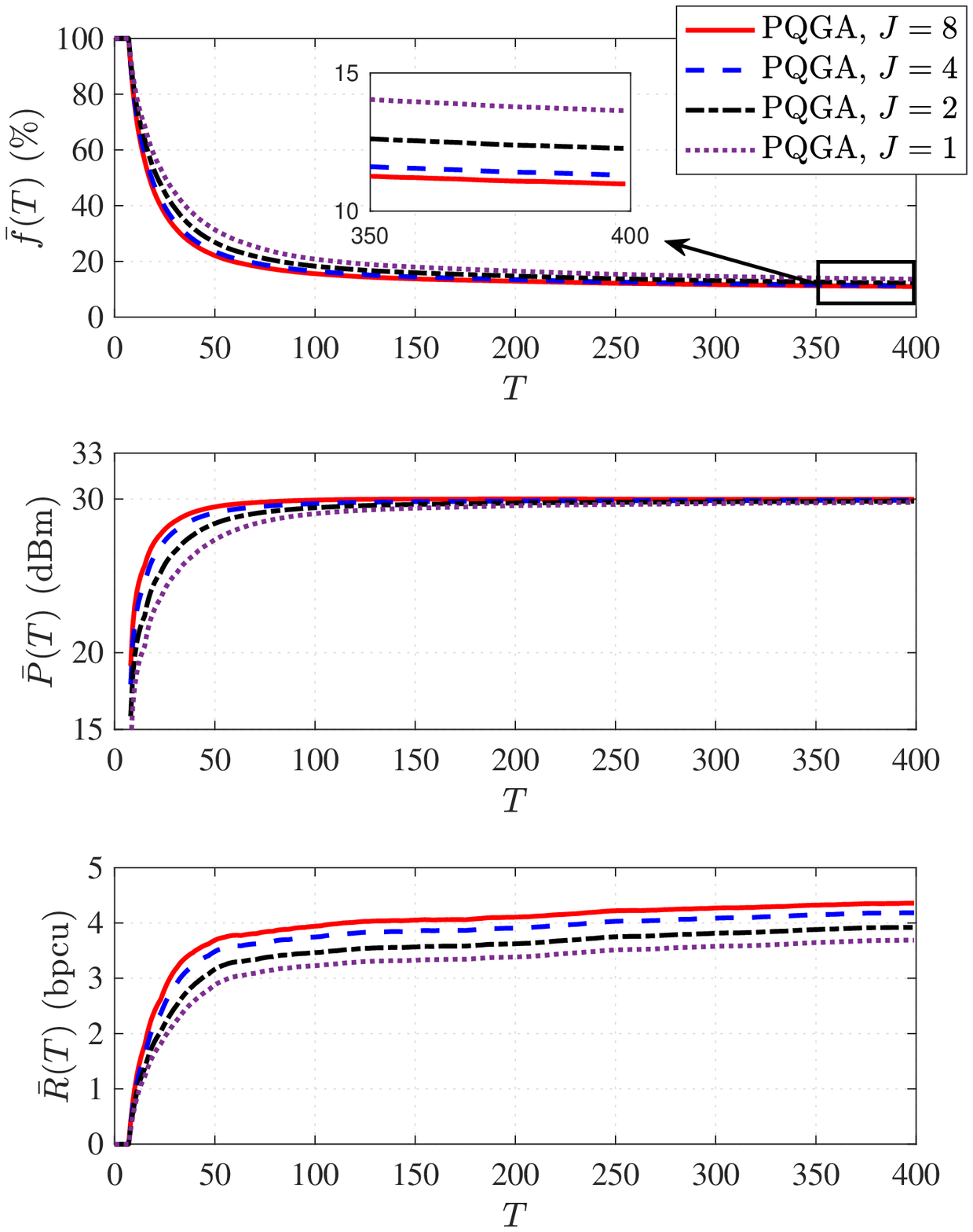}
\vspace{-4mm}
\caption{$\bar{f}$(T), $\bar{P}(T)$, and $\bar{R}(T)$ vs. $T$ with different
$J$ values.}\label{fig:J}
\end{minipage}
\vspace{-4mm}
\end{figure}

Fig.~\ref{fig:J} shows $\bar{f}(T)$, $\bar{P}(T)$ and $\bar{R}(T)$ versus
$T$ for different numbers of the aggregated gradient descent steps $J$, with
fixed update period $T_i=8$ and one CSI feedback. The system performance improves as $J$ increases, showing
the performance gain brought by performing multi-step gradient
descent. The impact of $J$ on the long-term transmit power $\bar{P}(T)$ is
small. We observe that the system performance almost stabilizes when $J=8$.
In the simulation results presented below, we set $J=8$ as default parameter
for PQGA.

\subsection{Performance Comparison}
\label{Sec:Performance Comparison}

For comparison, we consider the following method and performance benchmarks.
\begin{itemize}

\item \textit{Yu et al.:} The InP runs the online algorithm from \cite{LTC-HY}, which achieves the known best $\mathcal{O}(T^{\frac{1}{2}})$ static regret and $\mathcal{O}(1)$ constraint violation bounds. It has been demonstrated in \cite{LTC-HY} that this algorithm outperforms the ones in \cite{Trade} and \cite{LTC-Toff}.
Note that \cite{LTC-HY} considers only the standard OCO setting with per-time-slot
updates. In order to apply it to the periodic update scenario of our work,
we treat each update period of $T_i$ time slots as one super time slot. 
Furthermore, \cite{LTC-HY} considers only one gradient feedback at
each time slot. Therefore, if there are multiple CSI feedbacks in an update period,
we treat the averaged gradient as the gradient feedback.

\item \textit{Per-period optimal:} At the beginning of each update period
$i$, the InP has the CSI feedbacks from the current update period $i$,
and it implements $\mathbf{V}_{i}^\circ$ in~(\ref{EQ:xtopt}).

\item \textit{Delayed optimal:} At the beginning of each update period $i$,
 the InP collects the delayed CSI feedbacks from the previous update period
$i-1$, and it implements $\mathbf{V}_{i-1}^\circ$ in (\ref{EQ:xtopt}).

\item \textit{Offline fixed:} The InP has all the CSI feedbacks over $I$
update periods in hindsight, and implements $\mathbf{V}^\star$ in (\ref{EQ:xopt})
at each update period $i$.

\end{itemize}

\begin{figure}[!t]
\centering
\vspace{-2mm}
\begin{minipage}{0.495\textwidth}
\hspace{7.3mm}
\includegraphics[width=.845\linewidth,trim= 0 10 00 10,clip]{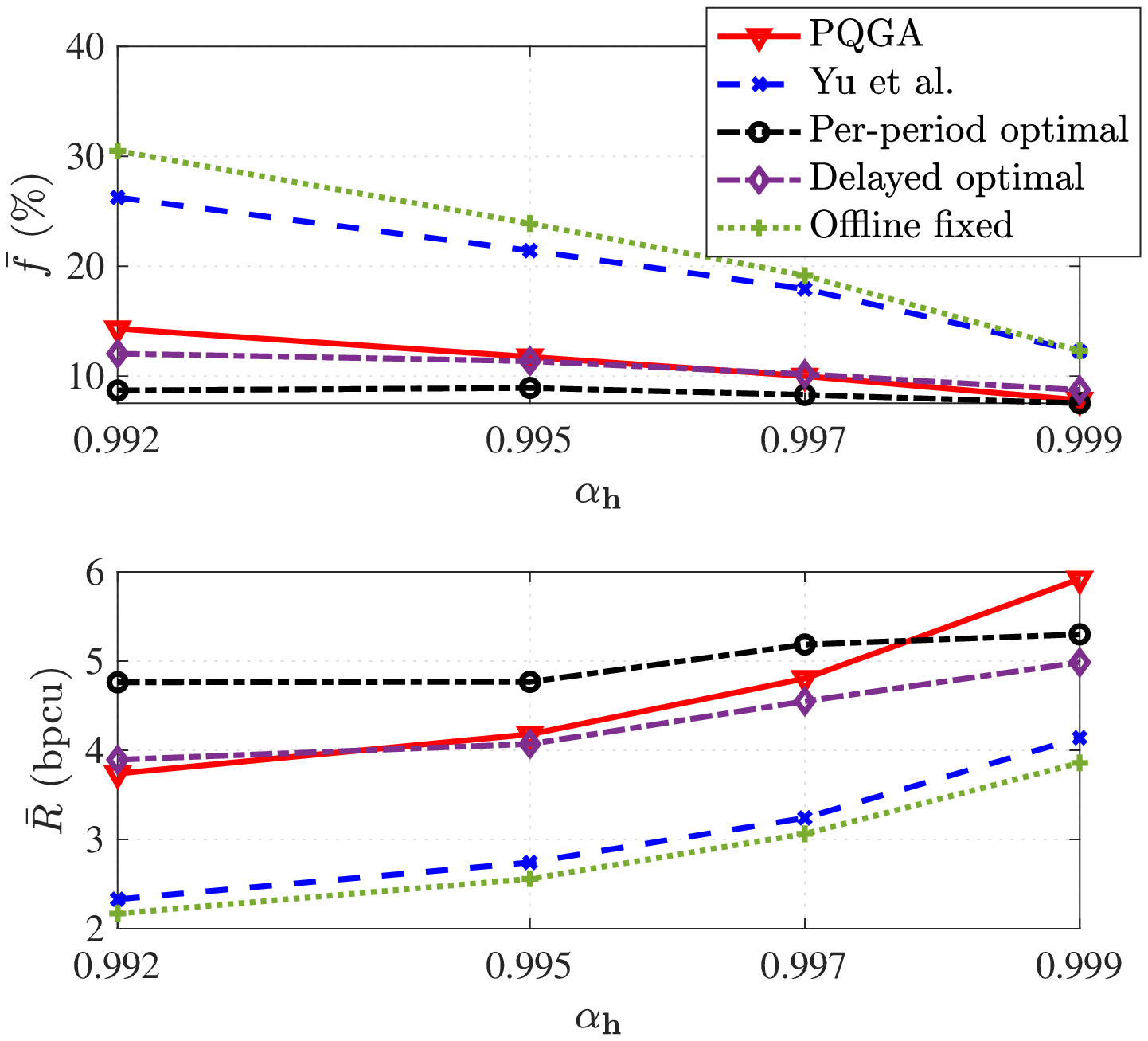}
\vspace{-4mm}
\caption{Performance comparison on $\bar{f}$ and $\bar{R}$ vs. $\alpha_{\mathbf{h}}$.}\label{fig:alpha}
\end{minipage}
\hfill
\begin{minipage}{0.495\textwidth}
\includegraphics[width=.845\linewidth,trim= 0 10 00 10,clip]{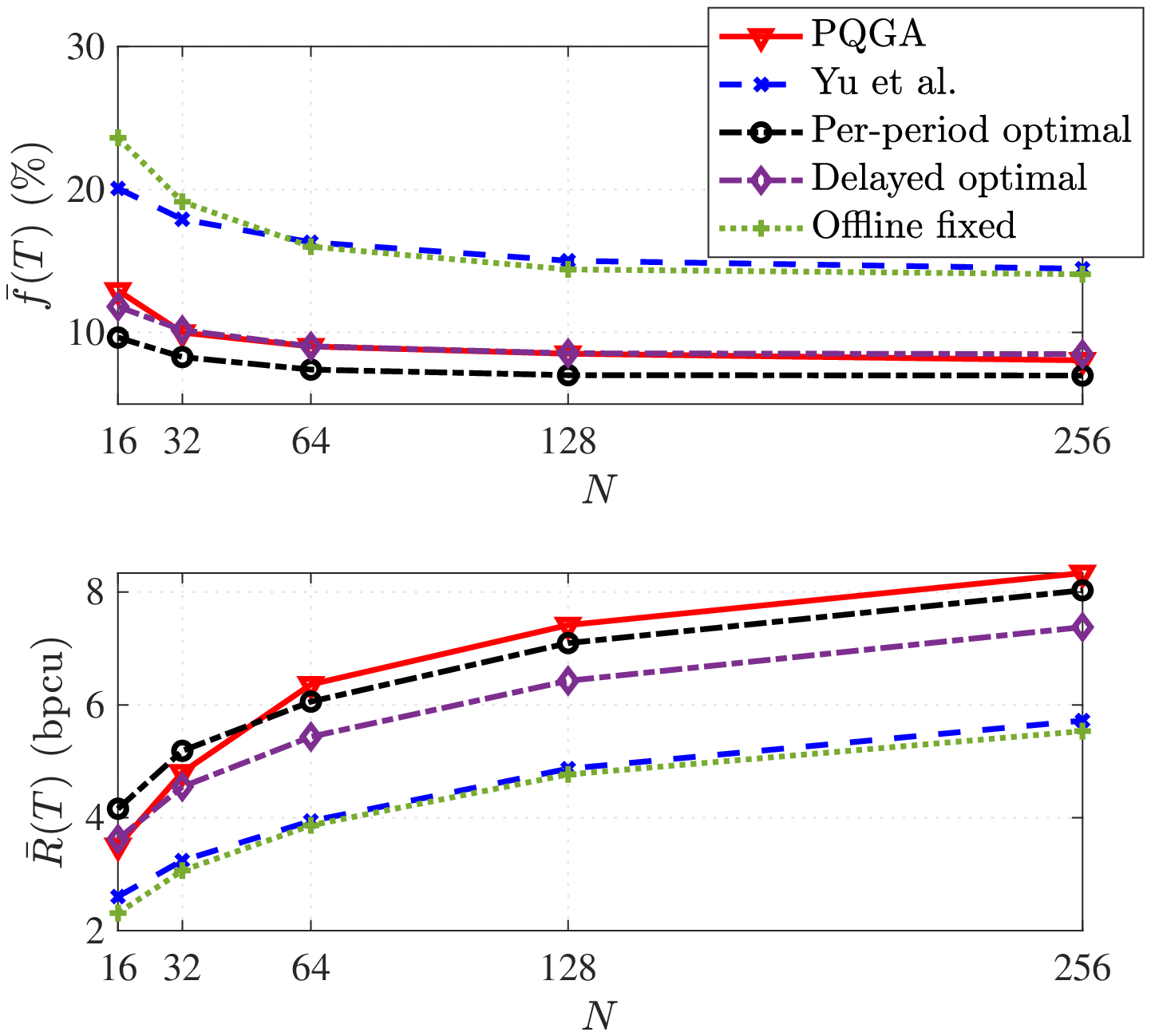}
\vspace{-4mm}
\caption{Performance comparison on $\bar{f}$ and $\bar{R}$ vs. $N$.}\label{fig:N}
\end{minipage}
\vspace{-4mm}
\end{figure}

We assume the update periods $\{T_i\}$ keep switching between $8$ and $4$ time slots. When $T_i=8$, there are two CSI feedbacks at the first and fifth time slot, \ie $S_i=2$. Otherwise, there is only one CSI feedback at the first time slot, \ie $T_i=4$ and $S_i=1$. Therefore, both the update periods $\{T_i\}$ and the numbers of CSI feedbacks $\{S_i\}$ are time varying.

In Fig.~\ref{fig:alpha}, we compare the steady-state precoding deviation
$\bar{f}$ and per-user rate $\bar{R}$ between PQGA, Yu et al., per-period optimal, delayed
optimal, and offline fixed, with different values of the channel correlation
coefficient $\alpha_{\mathbf{h}}$. There is a large performance gap between
the per-period optimal in (\ref{EQ:xtopt}) and the offline fixed in (\ref{EQ:xopt}),
indicating that the commonly used static benchmark for OCO may not be a meaningful
comparison target in practical time-varying systems.

As $\alpha_{\mathbf{h}}$ increases, the performance of PQGA improves, since
the channel states correlate more over time and the accumulated system variation
decreases. When $\alpha_\mathbf{h}>0.993$, which corresponds to pedestrian speed $2.5$~m/s, $\bar{f}$ yielded by
PQGA becomes smaller than the one produced by the delayed optimal.  Note that the per-period optimal has the current CSI and can be shown to have a semi-closed-form solution. In contrast, PQGA only uses the delayed CSI and has a closed-form solution with lower computational complexity. As $\alpha_{\mathbf{h}}$ increases, $\bar{f}$ yielded by PQGA can approach to the one yielded by the per-period optimal. Furthermore, it is more robust to channel variation compared
with the one yielded by Yu et al. under the periodic update setting. Note that the per-user rate $\bar{R}$ is not our precoding optimization objective, and thus is not directly related to $\bar{f}$. Furthermore, the per-period optimal needs to satisfy the long-term transmit power limit at each update period. As a side benefit, PQGA achieves a higher $\bar{R}$ than the per-period optimal when $\alpha_{\mathbf{h}}$ is large.

With the same setting, Fig.~\ref{fig:N} shows the impact of the number of antennas $N$ on the performance
of PQGA and the benchmarks. As $N$ increases, the InP has more freedom in
choosing the antennas for downlink beamforming to mitigate the inter-SP interference,
and thus the deviation from the virtualization demand $\bar{f}$ decreases.
As $N$ increases, $\bar{f}$ yielded by PQGA becomes smaller than the delayed
optimal and approaches to the per-period optimal. Furthermore,
the per-user rate $\bar{R}$ of PQGA is higher than the one yielded by
the per-period optimal when $N$ is large. PQGA substantially outperforms Yu et al. in a wide range of $N$ values.


\section{Conclusions}
\label{Sec:Conclusions}

This paper considers a new constrained OCO problem with periodic updates,
where the gradient feedbacks may be partly missing over some time slots and
the online decisions are updated over intervals that last for multiple time
slots. We present an efficient algorithm termed PQGA, which uses periodic
queues together with gradient aggregation to handle the possibly time-varying
delay between decision epoches. Our analysis considers the impact of the
constraint penalty structure and possibly multi-step aggregated gradient descent on the performance guarantees of PQGA, to provide bounds on its dynamic regret, static regret, and constraint violation. We apply PQGA to online network virtualization in massive MIMO systems. In addition to the benefits in terms of regret and constraint violation bounds, our simulation results further demonstrate the effectiveness of PQGA in terms of average system performance.

\bibliographystyle{IEEEtran}
\bibliography{References}

\begin{thebibliography}{10}
\providecommand{\url}[1]{#1}
\csname url@samestyle\endcsname
\providecommand{\newblock}{\relax}
\providecommand{\bibinfo}[2]{#2}
\providecommand{\BIBentrySTDinterwordspacing}{\spaceskip=0pt\relax}
\providecommand{\BIBentryALTinterwordstretchfactor}{4}
\providecommand{\BIBentryALTinterwordspacing}{\spaceskip=\fontdimen2\font plus
\BIBentryALTinterwordstretchfactor\fontdimen3\font minus
  \fontdimen4\font\relax}
\providecommand{\BIBforeignlanguage}[2]{{%
\expandafter\ifx\csname l@#1\endcsname\relax
\typeout{** WARNING: IEEEtran.bst: No hyphenation pattern has been}%
\typeout{** loaded for the language `#1'. Using the pattern for}%
\typeout{** the default language instead.}%
\else
\language=\csname l@#1\endcsname
\fi
#2}}
\providecommand{\BIBdecl}{\relax}
\BIBdecl

\bibitem{SPAWC20}
J.~{Wang}, B.~{Liang}, M.~{Dong}, and G.~{Boudreau}, ``Online {MIMO} wireless
  network virtualization over time-varying channels with periodic updates,'' in
  \emph{Proc. IEEE Intel. Workshop on Signal Process. Advances in Wireless
  Commun. (SPAWC)}, 2020.

\bibitem{OL}
N.~{Cesa-Bianchi} and G.~Lugosi, \emph{Prediction, Learning, and Games}.\hskip
  1em plus 0.5em minus 0.4em\relax Cambridge University Press, 2006.

\bibitem{OLBK}
S.~Shalev-Shwartz, ``Online learning and online convex optimization,''
  \emph{Found. Trends Mach. Learn.}, vol.~4, pp. 107--194, Feb. 2012.

\bibitem{UW}
P.~{Mertikopoulos} and A.~L. {Moustakas}, ``Learning in an uncertain world:
  {MIMO} covariance matrix optimization with imperfect feedback,'' \emph{IEEE
  Trans. Signal Process.}, vol.~64, pp. 5--18, Jan. 2016.

\bibitem{DC}
M.~{Lin}, A.~{Wierman}, L.~L.~H. {Andrew}, and E.~{Thereska}, ``Dynamic
  right-sizing for power-proportional data centers,'' \emph{IEEE/ACM Trans.
  Netw.}, vol.~21, pp. 1378--1391, Oct. 2013.

\bibitem{SG1}
Y.~{Zhang}, M.~H. {Hajiesmaili}, S.~{Cai}, M.~{Chen}, and Q.~{Zhu},
  ``Peak-aware online economic dispatching for microgrids,'' \emph{IEEE Trans.
  Smart Grid}, vol.~9, pp. 323--335, Jan. 2018.

\bibitem{Zinkevich}
M.~Zinkevich, ``Online convex programming and generalized infinitesimal
  gradient ascent,'' in \emph{Proc. Intel. Conf. Mach. Learn. (ICML)}, 2003.

\bibitem{NR}
\vspace{0mm}{3}GPP {TS}38.300, ``3rd {G}eneration {P}artnership {P}roject
  {T}echnical {S}pecification {G}roup {R}adio {A}ccess {N}etwork; {NR}; {NR}
  and {NG-RAN} {O}verall {D}escription; {S}tatge 2 ({R}elease 15).''

\bibitem{MEC}
B.~Liang, ``Mobile edge computing,'' in \emph{Key Technologies for 5G Wireless
  Systems}.\hskip 1em plus 0.5em minus 0.4em\relax V. W. S. {Wong}, R.
  {Schober}, D. W. K. Ng, and L.-C. {Wang}, Eds., Cambridge University Press,
  2017.

\bibitem{OT}
E.~{Hazan}, A.~{Agarwal}, and S.~{Kale}, ``Logarithmic regret algorithms for
  online convex optimization,'' \emph{Mach. Learn.}, vol.~69, pp. 169--192,
  2007.

\bibitem{slow}
J.~Langford, A.~J. Smola, and M.~Zinkevich, ``Slow learners are fast,'' in
  \emph{Proc. Adv. Neural Info. Proc. Sys. (NIPS)}, 2009.

\bibitem{AD}
K.~Quanrud and D.~Khashabi, ``Online learning with adversarial delays,'' in
  \emph{Proc. Adv. Neural Info. Proc. Sys. (NIPS)}, 2015.

\bibitem{E.C.Hall}
E.~C. {Hall} and R.~M. {Willett}, ``Online convex optimization in dynamic
  environments,'' \emph{IEEE J. Sel. Topics Signal Process.}, vol.~9, pp.
  647--662, Jun. 2015.

\bibitem{CDC}
A.~Jadbabaie, A.~Rakhlin, S.~Shahrampour, and K.~Sridharan, ``Online
  optimization : Competing with dynamic comparators,'' in \emph{Proc. Intel.
  Conf. Artif. Intell. Statist. (AISTATS)}, 2015.

\bibitem{Mokhtari16}
A.~{Mokhtari}, S.~{Shahrampour}, A.~{Jababaie}, and A.~{Ribeiro}, ``Online
  optimization in dynamic environments: Improved regret rates for strongly
  convex problems,'' in \emph{Proc. IEEE Conf. Decision Control (CDC)}, 2016.

\bibitem{L.Zhang17}
L.~{Zhang}, T.~{Yang}, J.~{Yi}, R.~{Jin}, and Z.~{Zhou}, ``Improved dynamic
  regret for non-degenerate functions,'' in \emph{Proc. Adv. Neural Info. Proc.
  Sys. (NIPS)}, 2017.

\bibitem{Dixit19}
R.~{Dixit}, A.~S. {Bedi}, R.~{Tripathi}, and K.~{Rajawat}, ``Online learning
  with inexact proximal online gradient descent algorithms,'' \emph{IEEE Trans.
  Signal Process.}, vol.~67, pp. 1338--1352, 2019.

\bibitem{Trade}
M.~Mahdavi, R.~Jin, and T.~Yang, ``Trading regret for efficiency: Online convex
  optimization with long term constraints,'' \emph{J. Mach. Learn. Res.},
  vol.~13, pp. 2503--2528, Sep. 2012.

\bibitem{LTC-Toff}
R.~Jenatton, J.~Huang, and C.~Archambeau, ``Adaptive algorithms for online
  convex optimization with long-term constraints,'' in \emph{Proc. Intel. Conf.
  Mach. Learn. (ICML)}, 2016.

\bibitem{X.CaoTau}
X.~Cao, J.~Zhang, and H.~V. Poor, ``Impact of delays on constrained online
  convex optimization,'' in \emph{Proc. Asilomar Conf. Signal Sys. Comput.},
  2019.

\bibitem{T.Chen}
T.~{Chen}, Q.~{Ling}, and G.~B. {Giannakis}, ``An online convex optimization
  approach to proactive network resource allocation,'' \emph{IEEE Trans.
  Signal. Process.}, vol.~65, pp. 6350--6364, Dec. 2017.

\bibitem{LTC-HY}
H.~Yu and M.~J. Neely, ``A low complexity algorithm with ${O}(\sqrt{T})$ regret
  and ${O}(1)$ constraint violations for online convex optimization with long
  term constraints,'' \emph{J. Mach. Learn. Res.}, vol.~21, pp. 1--24, Feb.
  2020.

\bibitem{Yu-SC}
H.~{Yu}, M.~J. {Neely}, and X.~{Wei}, ``Online convex optimization with
  stochastic constraints,'' in \emph{Proc. Adv. Neural Info. Proc. Sys.
  (NIPS)}, 2017.

\bibitem{X.Cao}
X.~{Cao}, J.~{Zhang}, and H.~V. {Poor}, ``A virtual-queue-based algorithm for
  constrained online convex optimization with applications to data center
  resource allocation,'' \emph{IEEE J. Sel. Topics Signal Process.}, vol.~12,
  pp. 703--716, Aug. 2018.

\bibitem{INFOCOM21}
J.~{Wang}, B.~{Liang}, M.~{Dong}, G.~{Boudreau}, and H.~{Abou-Zeid},
  ``Delay-tolerant constrained {OCO} with application to network resource
  allocation,'' in \emph{Proc. IEEE Conf. Comput. Commun. (INFOCOM)}, 2021.

\bibitem{Neely}
M.~J. Neely, \emph{Stochastic Network Optimization with Application on
  Communication and Queueing Systems}.\hskip 1em plus 0.5em minus 0.4em\relax
  Morgan \& Claypool, 2010.

\bibitem{Renewal}
M.~J. {Neely}, ``Dynamic optimization and learning for renewal systems,''
  \emph{IEEE Trans. Automat. Contr.}, vol.~58, pp. 32--46, 2013.

\bibitem{Lotfinezhad10}
M.~{Lotfinezhad}, B.~{Liang}, and E.~S. {Sousa}, ``Optimal control of
  constrained cognitive radio networks with dynamic population size,'' in
  \emph{Proc. IEEE Conf. Comput. Commun. (INFOCOM)}, 2010.

\bibitem{H.Yu}
H.~{Yu} and M.~J. {Neely}, ``Dynamic transmit covariance design in {MIMO}
  fading systems with unknown channel distributions and inaccurate channel
  state information,'' \emph{IEEE Trans. Wireless Commun.}, vol.~16, pp.
  3996--4008, Jun. 2017.

\bibitem{Wang20}
J.~{Wang}, M.~{Dong}, B.~{Liang}, and G.~{Boudreau}, ``Online precoding design
  for downlink {MIMO} wireless network virtualization with imperfect {CSI},''
  in \emph{Proc. IEEE Conf. Comput. Commun. (INFOCOM)}, 2020.

\bibitem{Gap}
O.~{Besbes}, Y.~{Gur}, and A.~{Zeevi}, ``Non-stationary stochastic
  optimization,'' \emph{Oper. Res.}, vol.~63, pp. 1227--1244, Sep. 2015.

\bibitem{rp}
V.~Jumba, S.~Parsaeefard, M.~Derakhshani, and T.~Le-Ngoc, ``Resource
  provisioning in wireless virtualized networks via massive-{MIMO},''
  \emph{IEEE Wireless Commun. Lett.}, vol.~4, pp. 237--240, Jun. 2015.

\bibitem{EE17}
Z.~Chang, Z.~Han, and T.~Ristaniemi, ``Energy efficient optimization for
  wireless virtualized small cell networks with large-scale multiple antenna,''
  \emph{IEEE Trans. Commun.}, vol.~65, pp. 1696--1707, Apr. 2017.

\bibitem{CRAN}
S.~Parsaeefard, R.~Dawadi, M.~Derakhshani, T.~Le-Ngoc, and M.~Baghani,
  ``Dynamic resource allocation for virtualized wireless networks in
  massive-{MIMO}-aided and fronthaul-limited {C-RAN},'' \emph{IEEE Trans. Veh.
  Technol.}, vol.~66, pp. 9512--9520, Oct. 2017.

\bibitem{WNVNOMA}
D.~{Tweed} and T.~{Le-Ngoc}, ``Dynamic resource allocation for uplink {MIMO}
  {NOMA} {VWN} with imperfect {SIC},'' in \emph{Proc. IEEE Int. Conf. Commun.
  (ICC)}, 2018.

\bibitem{AA}
Y.~{Liu}, M.~{Derakhshani}, S.~{Parsaeefard}, S.~{Lambotharan}, and K.~{Wong},
  ``Antenna allocation and pricing in virtualized massive {MIMO} networks via
  {S}tackelberg game,'' \emph{IEEE Trans. Commun.}, vol.~66, pp. 5220--5234,
  Nov. 2018.

\bibitem{V5G}
K.~Zhu and E.~Hossain, ``Virtualization of 5{G} cellular networks as a
  hierarchical combinatorial auction,'' \emph{IEEE Trans. Mobile Comput.},
  vol.~15, pp. 2640--2654, Oct. 2016.

\bibitem{Globecom19}
J.~Wang, M.~Dong, B.~Liang, and G.~Boudreau, ``Online downlink {MIMO} wireless
  network virtualization in fading environments,'' in \emph{Proc. IEEE Global
  Telecommun. Conf. (GLOBECOM)}, 2019.

\bibitem{Boyd}
S.~Boyd and L.~Vandenberghe, \emph{Convex Optimization}.\hskip 1em plus 0.5em
  minus 0.4em\relax Cambridge University Press, 2004.

\bibitem{LTEP}
H.~Holma and A.~Toskala, \emph{{WCDMA} for {UMTS} - {HSPA} evolution and
  {LTE}}.\hskip 1em plus 0.5em minus 0.4em\relax John Wiely \& Sons, 2010.

\end{thebibliography}

\end{document}